\documentclass{aa}
\usepackage[varg]{txfonts}

\usepackage{graphicx}
\usepackage{caption}
\usepackage{subcaption}

\usepackage{natbib}
\bibpunct{(}{)}{;}{a}{}{,}
\usepackage[citecolor=blue, linkcolor=blue, urlcolor = black, colorlinks = true]{hyperref}

\usepackage{longtable}
\usepackage{lscape}

\usepackage{verbatim}
\usepackage{makecell}
\usepackage{multirow}
\usepackage{booktabs}

\usepackage{xspace}
\usepackage{comment}
\usepackage{textcomp}
\usepackage[separate-uncertainty=true]{siunitx}
\usepackage[dvipsnames]{xcolor}
\usepackage[normalem]{ulem}

\newcommand*{\nstars}{2,497\ }

\begin{document}

\title{Evolution of the
near-core rotation frequency of \nstars
intermediate-mass stars from their dominant 
gravito-inertial mode\thanks{The
  full Tables B.1, B.2, B.3 are available at the CDS
via anonymous ftp to cdsarc.cds.unistra.fr (130.79.128.5)
or via https://cdsarc.cds.unistra.fr/viz-bin/cat/J/A+A/
???/A???}
}

\author{Conny Aerts\inst{\ref{KUL},\ref{Radboud},\ref{MPIA}}
\and Timothy Van Reeth\inst{\ref{KUL}} 
\and Joey S. G. Mombarg\inst{\ref{irap},\ref{KUL}, \ref{cea}}
\and Daniel Hey\inst{\ref{Hawaii}}
}
\institute{
Institute of Astronomy, KU Leuven, Celestijnenlaan 200D, B-3001
Leuven, Belgium \\
\email{conny.aerts@kuleuven.be} \label{KUL} 
\and Department of Astrophysics, IMAPP, Radboud University Nijmegen,
PO Box 9010, 6500 GL Nijmegen, The Netherlands\label{Radboud}
\and Max Planck Institute for Astronomy, K\"onigstuhl 17, 69117 Heidelberg, Germany\label{MPIA}
\and
IRAP, Universit\'e de Toulouse, CNRS, UPS, CNES, 14 avenue \'Edouard
Belin, F-31400 Toulouse,
France\label{irap}
\and
Universit\'e Paris-Saclay, Universit\'e de Paris, Sorbonne Paris
Cit\'e, CEA, CNRS,
AIM, 91191 Gif-sur-Yvette, France \label{cea}
\and
Institute for Astronomy, University of Hawai'i, Honolulu, HI 96822, USA\label{Hawaii}
}

\date{Received XXX / Accepted XXX}

	\abstract 
	    {The sparsely sampled time-series photometry from {\it
                Gaia\/} Data Release 3 (DR3) led to the discovery of
              more than 100,000 main-sequence non-radial
              pulsators. The majority of these were further
              scrutinised by uninterrupted high-cadence space
              photometry assembled by the Transiting Exoplanet Survey
              Satellite (TESS).}
	    {We combined {\it Gaia\/} DR3 and TESS photometric light
              curves to estimate the internal physical properties of
              \nstars gravity-mode pulsators.  We performed
              asteroseismic analyses with two major aims: 1) to
              measure the near-core rotation frequency and its
              evolution during the main sequence and 2) to estimate
              the mass, radius, evolutionary stage, and convective
              core mass from stellar modelling.}
	    {We relied on asteroseismic properties of {\it Kepler\/}
              $\gamma\,$Doradus and slowly pulsating B stars to derive
              the cyclic near-core rotation frequency, $f_{\rm rot}$,
              of the {\it Gaia}-discovered pulsators from their
              dominant prograde dipole gravito-inertial pulsation
              mode. Further, we investigated the impact of adding
              $f_{\rm rot}$ as an extra asteroseismic observable apart
              from the luminosity and effective temperature on the
              outcome of grid-based modelling from rotating stellar
              models. }
	    {We offer a recipe based on linear regression to deduce
              $f_{\rm rot}$ from the dominant gravito-inertial mode
              frequency. It is applicable to prograde
              dipole modes with an
              amplitude above 4\,mmag and occurring
              in the sub-inertial regime. By applying it to 2,497
              pulsators with such a mode, we have increased the sample of
              intermediate-mass dwarfs with such an asteroseismic
              observable by a factor of four. We used the estimate of
              $f_{\rm rot}$ to deduce spin parameters between two and six,
              while the sample's near-core rotation rates range from
              0.7\% to 25\% of the critical Keplerian rate. We used
              $f_{\rm rot}$, along with the {\it Gaia\/} effective
              temperature and luminosity to deduce the (convective
              core) mass, radius, and evolutionary stage from grid
              modelling based on rotating stellar models. We derived a
              decline of $f_{\rm rot}$ with a factor of two during the
              main-sequence evolution for this population of field
              stars, which covers a mass range from 1.3\,M$_\odot$ to
              7\,M$_\odot$. We found observational evidence for an
              increase in the radial order of excited gravity modes as
              the stars evolve. For 969 pulsators, we derived an upper
              limit of the radial differential rotation between the
              convective core boundary and the surface from {\it
                Gaia}'s {\tt vbroad} measurement and found values up to
              5.4.}
	    {Our recipe to deduce the near-core rotation frequency
              from the dominant prograde dipole gravito-inertial mode
              detected in the independent {\it Gaia\/} and TESS light
              curves is easy to use, facilitates applications to
              large samples of pulsators, and allows to map their angular
              momentum and evolutionary stage in the Milky Way. }
            

\titlerunning{Near-core rotation rates, (convective core) masses, radii, and evolutionary stages of 2497  gravity-mode pulsators}
\authorrunning{C.\ Aerts et al.}
\maketitle

\section{Introduction}

The internal rotation of stars continues to be one of the major
uncertain ingredients in the theory of stellar evolution
\citep{Maeder2009}. Modelling of the
dominant transport processes happening in the
interiors of stars relies on poorly calibrated expressions involving the
gradient of the angular rotation profile. To compute this gradient
throughout the evolution of the star, the so-called shellular
approximation introduced by \citet{Zahn1992} is often adopted. Relying
on this formulation, the profile of ${\rm d}\Omega (r)/{\rm d}r$ is
used to evaluate the element and angular momentum processes in the
stellar interior
\citep[e.g. ][]{Chaboyer1992,Heger2000,Maeder&Meynet2000}.

Over the past decade, asteroseismology \citep{Aerts2010} has been a
game changer in the study of stellar rotation. Progress on the
internal rotation of stars has relied on the detection and interpretation
of non-radial oscillation modes probing the deepest layers of stars.
\citet{Dupret2009} provided theoretical predictions on the capacity of
dipole mixed modes to probe the central regions of evolved low-mass
stars, while \citet{Miglio2008} and \citet{Bouabid2013} described the
properties of gravity and gravito-inertial modes in young
intermediate-mass dwarfs. These theoretical studies have been turned into
practical tools thanks to space asteroseismology \citep[][for recent
  overviews]
{HekkerJCD2017,GarciaBallot2019,Corsico2019,Aerts2021,Kurtz2022}.

Accurate values of the angular core or near-core rotation frequencies
for thousands of stars have been deduced. This was achieved from
dipole mixed modes in subgiants and red giants \citep[measuring the
  core value of the rotation, $\Omega_{\rm core}$,
  see][]{Beck2011,Bedding2011,Beck2012,Deheuvels2012,Mosser2012,Mosser2014,Mosser2015,Gehan2018}
and from dipole gravity or gravito-inertial modes in dwarfs
\citep[measuring the near-core value, $\Omega_{\rm near-core}$, just
  outside the convective core, see
][]{Degroote2010,Papics2012,Papics2014,Kurtz2014,Saio2015,VanReeth2015,VanReeth2016,Ouazzani2017,Papics2017,Saio2018}.
The diagnostic observables used to deduce the measurements of the
internal rotation can be rotationally split multiplets, or period
spacing patterns of mixed modes or gravity modes affected by the
Coriolis acceleration \citep[see ][for a review of the
  methodologies]{Aerts2019-araa}.  From the stars that also have a
measurement of the angular envelope rotation frequency ($\Omega_{\rm
  env}$), it was further established that strong (near-)core to
envelope coupling occurs, keeping the level of differential rotation
($\Omega_{\rm (near-)core}/\Omega_{\rm env}$) modest during the two
longest phases of stellar evolution
\citep{Deheuvels2015,Triana2015,VanReeth2018,GangLi2020,Ouazzani2020,Saio2021,GangLi2024}.
Further, it was established that the core rotation at the time of
central helium exhaustion is in agreement with the rotation of white
dwarfs \citep{Hermes2017,Aerts2019-araa}.  Moreover, $\Omega_{\rm
  (near-)core}/\Omega_{\rm env}$ increases strongly during hydrogen
shell burning, reaching values of up to 20
\citep{Deheuvels2014,DiMauro2016,Triana2017,Deheuvels2020,GangLi2024}.

Thanks to asteroseismology, our understanding of how angular momentum
transport and angular momentum loss slow down the (core) rotation
after exhaustion of the central hydrogen has improved dramatically
\citep{Fuller2019,Eggenberger2019a,Eggenberger2019b,Moyano2023a}. These
novel transport theories comply with the asteroseismic measurements of
the internal rotation for thousands of evolved stars of low and
intermediate mass.  For the longest phase of stellar evolution,
however, less than 100 dwarfs have both their 
$\Omega_{\rm near-core}$ and $\Omega_{\rm env}$,
or $\Omega_{\rm near-core}$ and $\Omega_{\rm surf}$
measured, where $\Omega_{\rm surf}$ is the
angular rotation frequency at the stellar surface
\citep{GangLi2020,Pedersen2021,Pedersen2022b}. Stars with these
observables available cover the mass range of $[1.3,9]\,$M$_\odot$ and span a
broad interval of measured near-core rotation rates \citep[up to about
  30\,$\mu$Hz,][]{Aerts2021}.  As a consequence, our understanding of
angular momentum transport during core hydrogen burning is still
limited, despite progress over the past five years
\citep{Fuller2019,Eggenberger2022,Betrisey2023,Moyano2023b,Moyano2024}.

Here, our aim is to enlarge the sample of dwarfs with an estimate of
the near-core rotation frequency and to study how it evolves in time
during the main-sequence phase. We wish to deduce ${\rm d} \Omega/{\rm
  d}t$ in the transition zone from the convective core to the
radiative envelope as stars evolve from the zero-age main sequence
(ZAMS) to the terminal-age main sequence (TAMS).  We achieve this by
relying on a sample of thousands of new gravity-mode pulsators
discovered from Data Release 3 (DR3) of the {\it Gaia\/} space mission
\citep{DeRidder2023,Aerts2023}.  \citet{HeyAerts2024} confirmed the
pulsational nature of these stars on the basis of independently
assembled light curves from the Transiting Exoplanet Survey Satellite
(TESS). The sample of gravity-mode pulsators with both {\it Gaia\/}
DR3 and TESS data in the public domain is 23 times larger than the
{\it Kepler\/} sample of dwarfs with measurements of $\Omega_{\rm
  near-core}$ \citep{MombargAerts2024}.

We worked towards our goal by relying on two well-established
properties of main-sequence gravity-mode pulsators deduced from {\it
  Kepler\/} space-based photometry: 1) The majority of such pulsators
reveal dominant prograde dipole modes
\citep{VanReeth2016,GangLi2020}; 2) the dominant mode in their
amplitude spectra reveals a clear shift towards higher frequency
according to their near-core rotation rate
\citep{Papics2017,AertsTkachenko2024}. In this work, we bring these two
properties together and come up with a regression formula for the
near-core rotation frequency from a measurement of the dominant
oscillation frequency of the {\it Kepler\/} gravity-mode pulsators
(Sect.\,2). We apply our recipe to the new sample of confirmed
gravity-mode pulsators discovered from {\it Gaia\/} and validated by
TESS (Sect.\,3).  We perform grid modelling based on the near-core
rotation frequency and the {\it Gaia\/} effective temperature and
luminosity as non-seismic observables and compare the outcome with
published evolutionary grid modelling without an asteroseismic
estimate of $\Omega_{\rm near-core}$ (Sect.\,4).  We end with
conclusions and the prospect of deriving an asteroseismic measurement
of the near-core rotation frequency for millions of dwarfs across the
Milky Way from future combined {\it Gaia\/} DR4 and TESS data
(Sect.\,5).

\section{A recipe for the near-core rotation frequency from a prograde dipole gravito-inertial mode}

{\it Gaia\/} DR3 led to the discovery of more than 100,000 new
candidate (non-)radial pulsators. Their dominant frequency have an
amplitude above about 4\,mmag \citep{DeRidder2023}. Among the new
non-radial pulsators, \citet{Aerts2023} selected those with an
amplitude below 35\,mmag and dominant frequency in the range
[0.7,3.2]\,d$^{-1}$. These two extra criteria were found to be
effective in selecting the $\gamma\,$Doradus ($\gamma\,$Dor) and
slowly pulsating B (SPB) star candidates among the overall sample
presented by \citet{DeRidder2023}. The global stellar parameters of
these 15,062 candidate gravity-mode pulsators were found to be in good
agreement with those of confirmed and well-studied {\it Kepler\/}
pulsators of these two kinds \citep{Aerts2023}.

\citet{HeyAerts2024} revisited the sample of more than 100,000
candidate main-sequence pulsators classified by
\citet{DeRidder2023}. They extracted light curves for almost 60,000 of
them from TESS Full Frame Images (FFIs) in the public domain. These
densely sampled TESS light curves allowed them to confirm the dominant
frequency found in the sparsely sampled {\it Gaia\/} DR3 time-series
photometry for the majority of them. \citet{HeyAerts2024} then
reclassified the pulsators with a TESS FFI light curve. They came up
with a cleaner sub-sample of gravity-mode pulsators among the
multiperiodic variables, as improvement with respect to the
\citet{Aerts2023} study.

Here, we consider a sub-sample of 15,692 pulsators classified by
\citet{HeyAerts2024} with a probability above 50\% to be a
$\gamma\,$Dor or SPB star or a hybrid pressure/gravity-mode pulsator.
\citet{MombargAerts2024} determined the mass, radius, and evolutionary
stage of these 15,692 pulsators from evolutionary grid modelling. This
was done by matching the stars' {\it Gaia\/} DR3 effective temperature
and luminosity with those predicted from a grid of rotating stellar
models calibrated by asteroseismology of {\it Kepler\/} gravity-mode
pulsators, covering the mass range from 1.3\,M$_\odot$ to
9\,M$_\odot$.

For our study, we rely on the dominant pulsation mode of these 15,692
stars found consistently in the {\it Gaia\/} and TESS light curves to
assess their cyclic near-core rotation frequency, denoted here for
simplicity as $f_{\rm rot}\!\equiv\Omega_{\rm near-core}/2\pi$. In
order to achieve this, we first recall some general properties of
well-studied {\it Kepler\/} gravity-mode pulsators.

\subsection{Basic properties of gravito-inertial modes in dwarfs}

Main-sequence pulsators of $\gamma\,$Dor or SPB type exhibit
low-frequency gravity modes of high radial-order, $n$, excited by
either the flux blocking mechanism or the $\kappa\,$mechanism
\citep{Pamyatnykh1999,Guzik2000,Bouabid2013,Xiong2016,Szewczuk2017}. The
majority of these pulsators experience moderate to fast rotation,
which brings the oscillation modes into the sub-inertial regime, where
both buoyancy and the Coriolis force act together as restoring forces
\citep{Aerts2019-araa,AertsTkachenko2024}. Such modes are called
gravito-inertial modes, while modes in slow rotators restored by
buoyancy alone are pure gravity modes.

In the asymptotic frequency limit, where the pulsation frequency in
the co-rotating frame, $f_{\rm co}$, is much smaller than the
Brunt-V\"ais\"al\"a frequency, $N$, and in the `Traditional
Approximation of Rotation' \citep[TAR;
  e.g.\ ][]{LeeSaio1987,Townsend2003}, we have
\begin{equation}
f_{\rm co} = \frac{\sqrt{\lambda_{km\nu}}}{\Pi_0\left(n + \alpha_g\right)},
\label{eq:tar-pattern}
\end{equation}
where $\lambda_{km\nu}$ is the eigenvalue of the Laplace tidal equation 
for a mode labelled as $(k,m)$ having
spin parameter $\nu = 2f_{\rm rot}/f_{\rm co}$, 
$\alpha_g$ is a phase term dependent on the boundaries of the
pulsation mode cavity denoted by the positions $r_1$ and $r_2$, and
the buoyancy radius is
\begin{equation}
\Pi_0 = \frac{2\pi^2}{\int_{r_1}^{r_2} \frac{N(r)}{r}\mathrm{d}r}.
\end{equation}
For non-rotating stars, gravity modes are labelled by indices
$(l,m,n)$, where $l$ and $|m|\leq l$ denote the degree and azimuthal
order of the spherical harmonic $Y_l^m$ describing the displacement
vector of the mode. In this work, we define prograde and retrograde
modes as having $m>0$ and $m<0$, respectively, while zonal modes have
$m=0$.  In order to encompass a single mode classification procedure
for rotating pulsators, \citet{LeeSaio1997} defined a suitable
labelling scheme based on one unique integer index $k$ because $l$ is
not suitable to capture all mode families. The value of $k$ is able to
represent all families of solutions to the Laplace tidal equation,
which describes the oscillations when the vertical component of the
Coriolis acceleration is taken into account, but not the horizontal
component. In this labelling scheme, zero or positive values of $k$
correspond to modes having a counterpart of the harmonic degree
$l=|m|+k$ in the limit of no rotation. On the other hand, negative $k$
values stand for retrograde modes that do not have any counterpart in
the non-rotating case \citep[see also][for more details on mode
  classification schemes in rotating pulsators]{Townsend2003}. Our
study is focused on stars with $k\geq 0$ modes. In particular, zonal
dipole gravity modes are labelled as $(k,m)=(1,0)$, tesseral
quadrupole retrograde gravity modes have $(k,m)=(1,-1)$, while
prograde sectoral dipole gravity or gravito-inertial modes have
$(k,m)=(0,1)$, all of which occur in
Fig.\,\ref{fig:Kepler-relation-frot} discussed below.

The condition that the mode frequencies must be well below $N$ is
fulfilled for $\gamma\,$Dor and SPB stars, as shown by \citet[][see
  their Fig.\,1]{Aerts2021-GIW}. Moreover, \citet{Aerts2017},
\citet{Aerts2021-GIW}, and \citet{Pedersen2022b} deduced the spin
parameters of the identified modes of $\gamma\,$Dor and SPB pulsators
and found $\nu\geq 1$ for almost all modes in the $\gamma\,$Dor stars
and for the majority of them in SPB stars, justifying the use of the
TAR in general terms for such pulsators \citep[see][for extensive
  discussions on the TAR criteria]{Mathis2013-LNP}.

In the limit of high radial order $n>\!>1$ (corresponding to the limit
of low frequencies), gravity or gravito-inertial modes of the same
$(k,m)$ and consecutive radial order $n$ obey a regular pattern. This
is predicted theoretically
\citep{Miglio2008,Bouabid2013,VanReeth2015a,Ouazzani2017} and observed
in photometry of single and binary pulsators assembled with the
Convection, Rotation, and exoplanetary Transits (CoRoT), {\it
  Kepler\/} or TESS space telescopes
\citep[e.g.\ ][]{Degroote2010,Papics2012,Papics2014,Kurtz2014,Saio2015,VanReeth2015,Keen2015,Papics2017,
  Szewczuk2018,Sekaran2020,Wu2020,GangLi2020,Szewczuk2021,Pedersen2021,Garcia2022,Kemp2024}.
These publications revealed the regular mode patterns detected in the
space photometry to allow for mode identification.  For the
$\gamma\,$Dor stars, most of the observed modes identified as
$(k,m)=(0,1)$ in these publications have radial order $n\in [10,100]$
\citep{GangLi2020}, while SPB stars typically have $(k,m)=(0,1)$ modes
with $n\in [10,50]$ excited \citep{Szewczuk2017}.

As elaborated by \citet{AertsTkachenko2024}, the Coriolis force is
more important than the the centrifugal force in the calculation of
the modes for $\gamma\,$Dor and SPB stars. This is due to the mode
energy being dominated by the region adjacent to the convective core,
where the rotational deformation due to the centrifugal force is
modest
\citep{Mathis2019,Henneco2021,Dhouib2021a,Dhouib2021b}. Moreover, the
horizontal components of the displacement vector of the modes are
typically an order of magnitude larger than the vertical ones
\citep{DeCatAerts2002}.  All of these properties together imply that
the TAR is well justified, unless the stars rotate close to their
critical rate. We discuss and evaluate this latter condition in
Sect.\,4.

\subsection{Identification of the dominant mode of the {\it Kepler\/} gravity-mode pulsators}

The combined {\it Kepler\/} gravity-mode pulsator samples analysed by
\citet{GangLi2020}, \citet{Pedersen2021}, and \citet{Pedersen2022a}
reveal period-spacing patterns of prograde dipole modes for 610 of the
634 stars in the combined sample ($96.2\%$). For 428 of these, that is
$67.5\%$ of the sample, the dominant mode was part of the detected
pattern. When only stars with dominant pulsation amplitudes larger
than 4\,mmag are considered among these {\it Kepler\/} pulsators,
which is the amplitude detection threshold for the discovery of the
{\it Gaia\/} DR3 gravity-mode pulsators, this fraction increases to
$84.7\%$.

Hence, in the remainder of this work we rely on the fact that the
large majority among the dominant frequencies of well-characterised
{\it Kepler\/} $\gamma\,$Dor and SPB stars correspond to a mode with
$(k,m) = (0,1)$.  While doing so, we keep in mind that about 15\% of
the dominant frequencies may belong to a mode with another
identification of $(k,m)$.

\subsection{\label{subsec:regr} A regression formula for the internal rotation frequency}

\begin{figure*}[h!]
 \includegraphics[width=\textwidth]{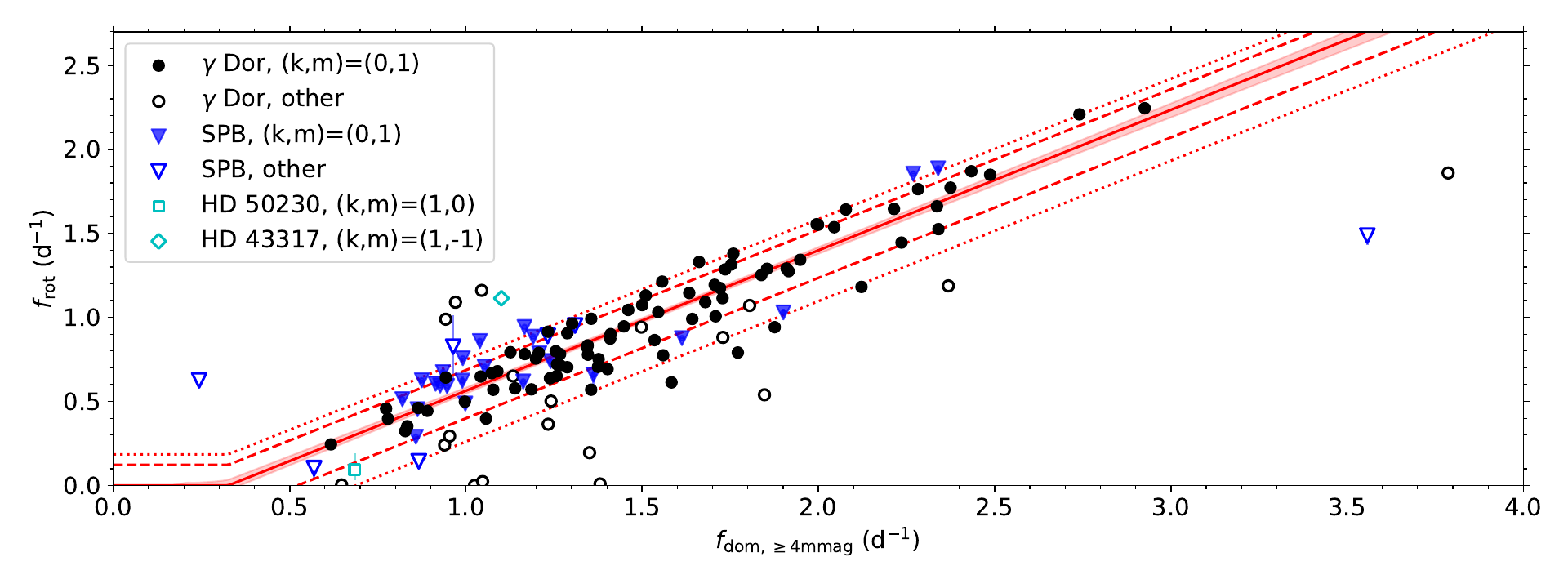}
 \caption{Near-core rotation frequency $f_{\rm rot}$ plotted as a
   function of dominant mode frequency for {\it Kepler\/} gravity-mode
   pulsators having an observed amplitude above 4\,mmag $(f_{{\rm
       dom,}\ \geq\,4\,{\rm mmag}})$ from the samples of
   \citet{GangLi2020} and \citet{Pedersen2021}.  Stars are drawn as
   $\gamma$\,Dor pulsator if they have $T_{\rm eff} < 8500\,\rm K$ and
   $\Pi_0 < 0.7\,$d and as SPB star otherwise.  When invisible, the
   error bars are smaller than the symbol sizes.  The linear relation
   (red line) given by Eq.~(\ref{eq:frot-relation}) was computed for
   the 105 stars having a dominant $(k,m)=(0,1)$ mode with observed
   frequency above 0.35\,d$^{-1}$ and with $\nu > 1$ (full
   symbols). The shaded area shows the 1-$\sigma$ uncertainty region
   for this relation.  The two pulsators indicated in cyan come from
   short CoRoT light curves \citep{Degroote2010,Papics2012}.  As the
   other open symbols, they were not used in the regression.  The
   dashed (dotted) lines mark the 1-$\sigma$ confidence interval for
   $f_{\rm rot}$ derived from the residuals with respect to the red
   line, excluding (including) stars with modes having $(k,m)\neq
   (0,1)$ or $\nu < 1$.}
 \label{fig:Kepler-relation-frot}
\end{figure*}
\begin{figure}[h!]
 \includegraphics[width=88mm]{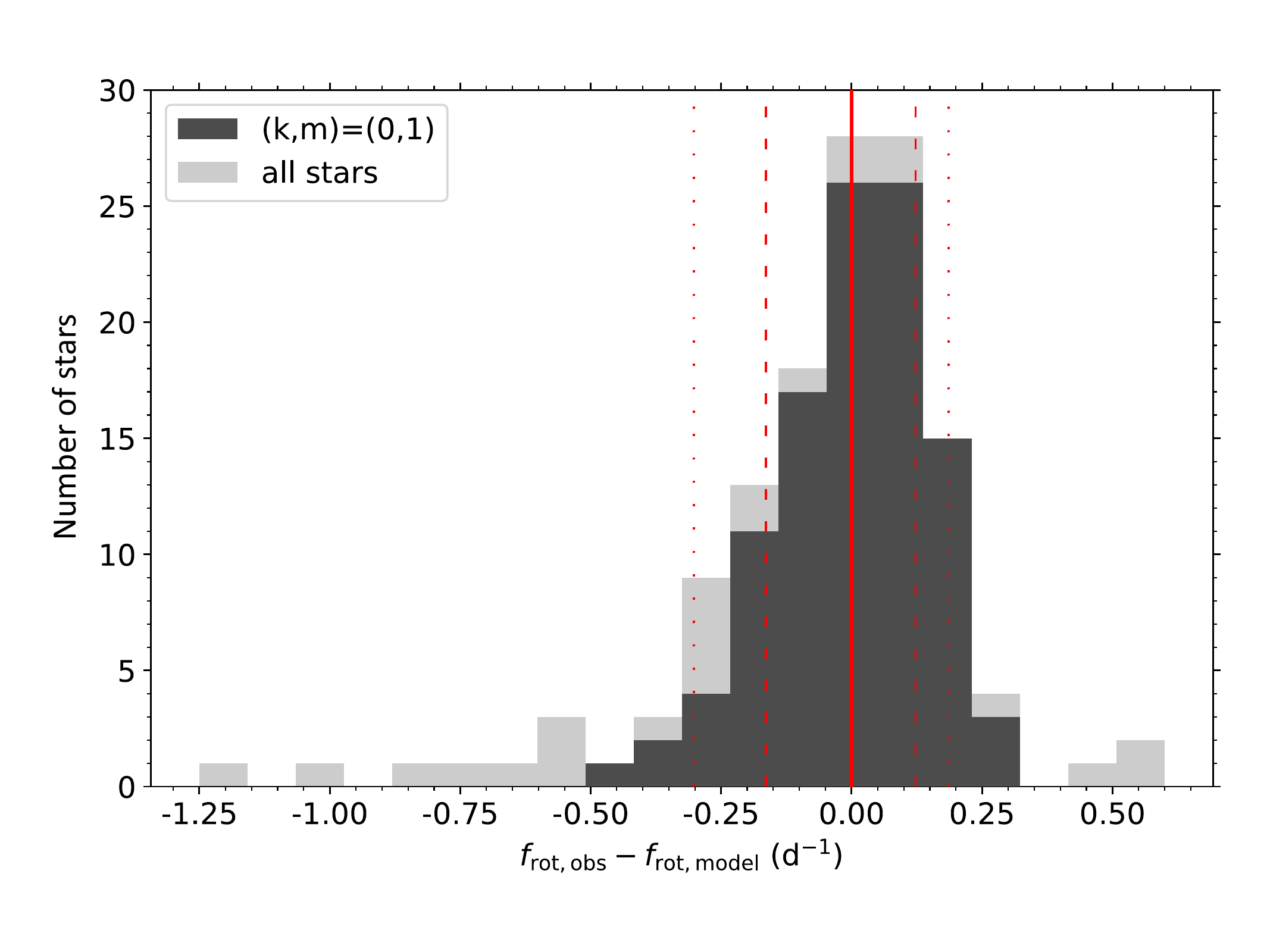}
 \caption{Residuals from the linear fit described by
   Eq.\,(\ref{eq:frot-relation}) for the selected {\em Kepler}
   pulsators used in Fig.\,\ref{fig:Kepler-relation-frot}. The
   distribution of modes with $(k,m)=(0,1)$, $\nu > 1$, and observed
   frequency above 0.35\,d$^{-1}$ is shown in dark grey, wile the
   distribution of all dominant modes with amplitude $\geq 4\,\rm
   mmag$ is shown in light grey. The full, dashed, and dotted lines
   indicate the same relation and confidence intervals as those in
   Fig.\,\ref{fig:Kepler-relation-frot}. \label{fig:Kepler-residuals-frot}}
\end{figure}

In the case of modes with $k = 0$ in moderate- to fast-rotating stars,
that is, with $\nu \gg 1$, theory predicts that $\lambda_{km\nu}
\approx m^2$ \citep[e.g. ][]{Townsend2003}. This leads to a
quasi-linear relation between the observed cyclic gravity-mode
frequencies $f_{nkm}$ and the near-core rotation frequency $f_{\rm
  rot}$:
\begin{equation}
f_{\rm nkm} = f_{\rm co} + mf_{\rm rot} \approx \frac{m}{\Pi_0\left(n + \alpha_g\right)} + mf_{\rm rot},
\label{fit}
\end{equation}
which has also been found observationally
\citep[e.g.][]{VanReeth2016,Papics2017,GangLi2020,Audenaert2022}.
Here, we wish to exploit this relationship in the context of the newly
discovered {\it Gaia\/} gravity-mode pulsators, whose observed
dominant cyclic mode frequencies, $f_{\rm dom}$, are consistent in the
DR3 and TESS data and which have amplitudes above 4\,mmag.  An earlier
similar idea was put forward by \citet{Sepulveda2022} and
\citet{Sepulveda2023} in the context of the exoplanet host stars
51\,Eridani and HR\,8799, which are both $\gamma\,$Dor
pulsators. These authors considered an averaged frequency value among
the dominant modes to infer the near-core rotation frequency in an
empirical way by considering the sample of \citet{GangLi2020}, without
relying on the theory-based Eq.\,(\ref{fit}) as we do here.

To quantify the observational relation between $f_{\rm rot}$ and
$f_{\rm dom}$ we fitted a linear model to the {\em Kepler} results found
for the homogeneously studied samples of $\gamma\,$Dor and SPB
pulsators from \citet{GangLi2020} and \citet{Pedersen2021}. In light
of the application to the {\it Gaia\/} discovered gravity-mode
pulsators, we only selected the {\it Kepler\/} pulsators with dominant
pulsation amplitude above 4\,mmag and identified as $(k,m) = (0,1)$,
that is only stars with a dominant dipole prograde mode.  Moreover, we
demanded their dominant mode to occur in the sub-inertial regime, where
$\nu > 1$. In this way, we found the following regression fit based on
the 100 $\gamma\,$Dor stars and five SPB stars with their dominant prograde
dipole mode in the sub-inertial regime, having an observed amplitude above
4\,mmag, and having been subjected to 
asteroseismic modelling from {\it Kepler\/}:
\begin{equation}
f_{\rm rot} = 0.836^{+0.023}_{-0.027} f_{\rm dom} - 0.272^{+0.041}_{-0.036}\ {\rm d}^{-1},
\label{eq:frot-relation}
\end{equation}
where all of $f_{\rm rot}, f_{\rm dom}$, and $1/\Pi_0$ are expressed
in d$^{-1}$.  This fit is illustrated by the full red line in
Fig.\,\ref{fig:Kepler-relation-frot}. The strong correlation between
$f_{\rm rot}$ and $f_{\rm dom}$ is reflected by the Pearson
correlation coefficient, which has a value of 0.942.

We applied a bootstrap analysis on the residuals of the fit in
Eq.\,(\ref{eq:frot-relation}) and found that the estimated rotation
frequency values have uncertainties $\sigma_{-}$ and $\sigma_{+}$ of
$-0.164\,{\rm d}^{-1}$ and $+0.123\,{\rm d}^{-1}$, respectively.  The
results for this confidence interval are graphically depicted in
Fig.\,\ref{fig:Kepler-residuals-frot} by the dashed lines.  If we also
took into account the residuals with respect to the regression line
for the $\sim20\%$ sample stars that have pulsations with other
geometries $(k,m)\neq (0,1)$ or in the super-inertial regime
$(\nu<1$), we found uncertainties $\sigma_{-}$ and $\sigma_{+}$ of
$-0.302\,{\rm d}^{-1}$ and $+0.186\,{\rm d}^{-1}$, respectively. This
confidence interval is delineated by the dotted lines in
Figs\,\ref{fig:Kepler-relation-frot} and
\ref{fig:Kepler-residuals-frot}.

For illustrative purposes, we also plot the two gravity-mode pulsators
having an identified dominant mode from their 5-months-long light
curve assembled by the CoRoT space telescope in
Fig.\,\ref{fig:Kepler-relation-frot}. Even if their frequency
precision is lower than the one deduced from 4-year duration {\it
  Kepler\/} light curves, they still occur in the vicinity of the {\it
  Kepler\/} pulsators with $(k,m)\neq (0,1)$ modes. None of these
stars (with open symbols) were used in the construction of the
regression formula in Eq.\,(\ref{eq:frot-relation}).

\subsection{The scaled buoyancy radius}
\begin{figure*}
 \includegraphics[width=\textwidth]{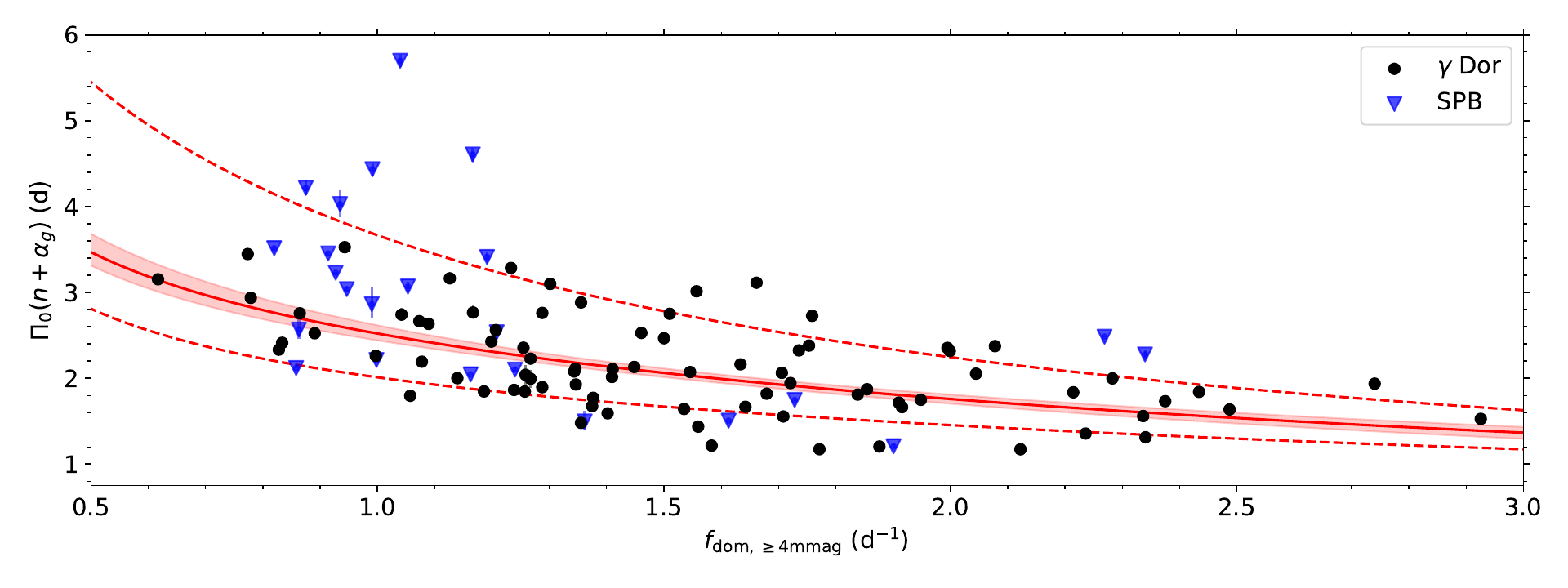}
 \caption{The quantity $\Pi_0(n+\alpha_g)$ plotted as a function of
   the dominant prograde dipole gravity-mode frequency with amplitude
   above 4\,mmag $(f_{{\rm dom,}\ \geq\,4\,{\rm mmag}})$ for 105
   $\gamma\,$Dor and SPB stars from \citet{GangLi2020} and
   \citet{Pedersen2021}. The symbols have the same meaning as in
   Fig.\,\ref{fig:Kepler-relation-frot}. The relation given by
   Eq.~(\ref{fit}) is shown by the red line, while the shaded area
   shows the 1-$\sigma$ uncertainty region. The dashed lines mark the
   1-$\sigma$ confidence region, propagated forward from the
   confidence regions of the $f_{\rm rot}$
   values. \label{fig:Kepler-relation-Pi0n}}
\end{figure*}

For the observed dominant prograde dipole modes in the {\it Kepler\/}
sample, we used Eq.\,(\ref{eq:tar-pattern}) to derive
$\Pi_0\left(n+\alpha_g\right)$, where we used the mode frequencies and
the determined rotation frequencies $f_{\rm rot}$ as input. The
results of this analysis are shown in
Fig.\,\ref{fig:Kepler-relation-Pi0n}. Here, the derived 1-$\sigma$
confidence intervals were propagated from the confidence intervals of
$f_{\rm rot}$, derived using stratified residual bootstrapping
\citep{DavisonHinkley1997}.  The relative scatter of the
$\Pi_0\left(n+\alpha_g\right)$ values is larger than of the $f_{\rm
  rot}$ values, as reflected by a lower Pearson correlation
coefficient of 0.542.

While most {\em Kepler} sample stars are in agreement with the
confidence regions, a few outliers with measured mode frequencies
between 1 and 1.2$\,\rm d^{-1}$ occur.  These stars from
\citet{GangLi2020} are SPB stars based on their $T_{\rm eff}$ and our
definition. They have anomalously high $\Pi_0$ values, which may
result from an overestimated $f_{\rm rot}$ value propagating into the
derived high $\Pi_0\left(n+\alpha_g\right)$ values above 4.5\,d.

\section{Application to a {\it Gaia} sample}

\begin{figure}
    \includegraphics[width=\linewidth]{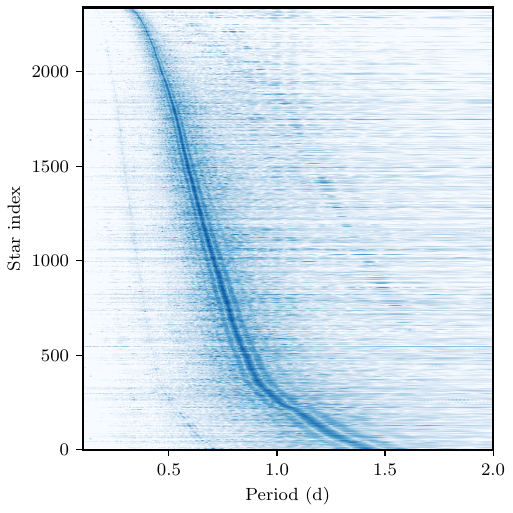}
    \caption{Stacked periodogram of the \nstars sample stars
      calculated for the TESS light curves from
      \citet{HeyAerts2024}. Each row corresponds to the amplitude
      spectrum of one star in the gravity-mode sub-sample treated
      here, sorted by the dominant mode period. For all these \nstars
      pulsators, the dominant mode period occurs in the ridge of
      prograde dipole gravito-inertial $(k,m)=(0,1)$ modes.
    \label{fig:Dan}}
\end{figure}

\citet[][their Fig.\,6]{HeyAerts2024} found a distinct ridge in the
stacked amplitude periodograms of the gravity-mode pulsators,
characteristic for prograde dipole modes as also found for the {\it
  Kepler\/} $\gamma\,$Dor pulsators by \citet{GangLi2020}. Among the
15,692 {\it Gaia\/} DR3 gravity-mode pulsators modelled by
\citet{MombargAerts2024}, we defined a sub-sample with the aim to
select a maximum number of stars with a high probability of having a
dominant prograde dipole mode in the TESS FFI light curve.

We required that {\em (i)} the amplitudes and frequencies of the
dominant pulsations in the {\em Gaia} DR3 light curves are $\geq4\,\rm
mmag$ and $\leq 3.2\,\rm d^{-1}$, respectively \citep{Aerts2023}, {\em
  (ii)} they were classified by \citet{HeyAerts2024} as $\gamma\,$Dor
or SPB star or else as hybrid gravity/pressure mode pulsator, in all
of these three cases with a probability above 50\,\%, {\em (iii)} the
dominant frequencies measured from the {\em Gaia} and TESS data differ
less than $0.1\,\rm d^{-1}$ \citep{HeyAerts2024}, {\em (iv)}
$\log\left(L/L_\odot\right)\geq 0$, and {\em (v)} $G_{Gaia} \leq
13\,\rm mag$. These stringent constraints delivered a sub-sample of
\nstars pulsators among those with parameters from evolutionary
modelling by \citet{MombargAerts2024}, notably their mass $M_\star$,
convective core mass $m_{\rm cc}$, radius $R_\star$, and evolutionary
stage defined as the ratio of the hydrogen mass fraction remaining in
the convective core compared to the initial value at birth,
$X_c/X_{\rm ini}$.

Following \citet{HeyAerts2024}, Fig.\,\ref{fig:Dan} shows the stacked
periodograms of these \nstars gravity-mode pulsators based on their
TESS FFI light curves. A prominent ridge occurs, very similar to the
one found for stars with dominant identified $(k,m)=(0,1)$ modes from
{\it Kepler\/} light curves by \citet{GangLi2020}. This lends support
to our assumption that these {\it Gaia\/} gravity-mode pulsators have
a dominant prograde dipole gravito-inertial mode. In the remainder of
this work we reasonably assume that all the \nstars stars in our
current sub-sample with periodogram shown in Fig.\,\ref{fig:Dan}
fulfil the mode identification $(k,m) = (0,1)$.

We relied on the relations derived from the {\em Kepler}
gravity-mode pulsators deduced in the previous Section to infer the
near-core rotation rates $f_{\rm rot}$ and scaled buoyancy radius
$\Pi_0\left(n+\alpha_g\right)$ for the \nstars stars in the {\em Gaia}
DR3 sub-sample. Uncertainties for these two inferred quantities were
estimated using the confidence intervals defined in
Sect.\,\ref{subsec:regr} and taking the frequency differences
$|f_{{\rm dom}, Gaia} - f_{{\rm dom, TESS}}|$ as an additional
systematic observational error for $f_{\rm dom}$, which we then
propagated forward using the chain rule.
\begin{figure}[th!]
\rotatebox{0}{\resizebox{8.8cm}{!}{\includegraphics{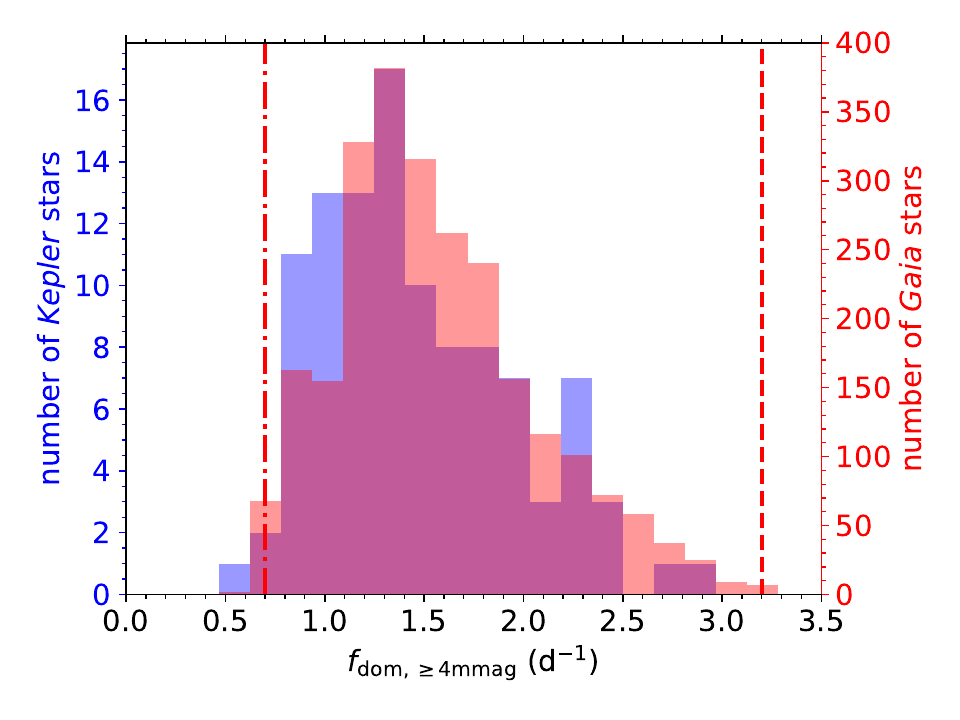}}}\\[-0.3cm]
\rotatebox{0}{\resizebox{8.8cm}{!}{\includegraphics{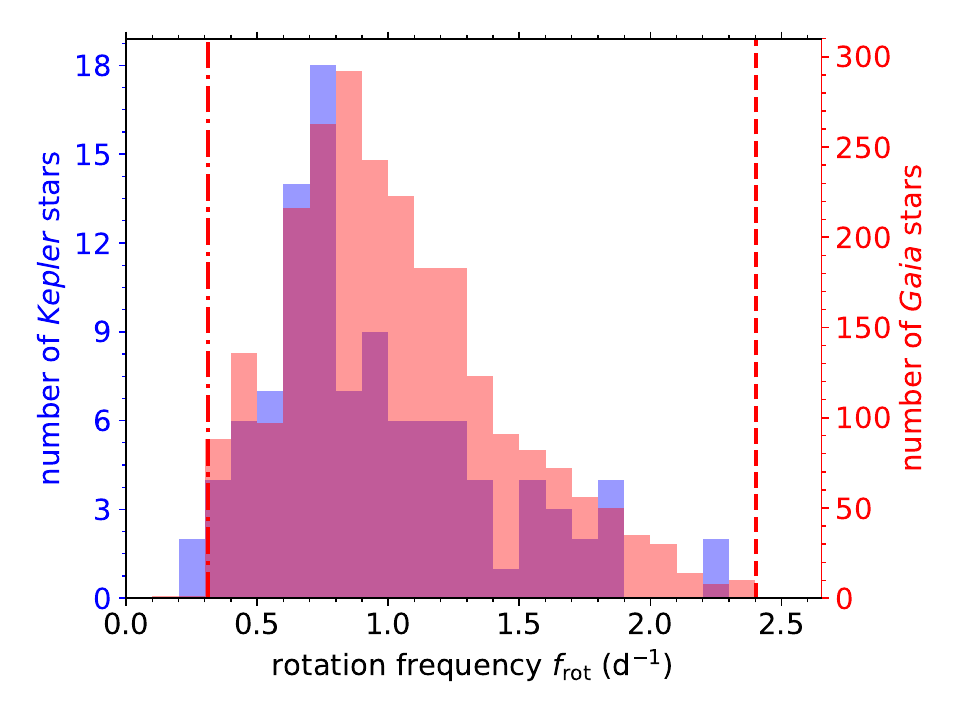}}}\\[-0.3cm]
\rotatebox{0}{\resizebox{8.8cm}{!}{\includegraphics{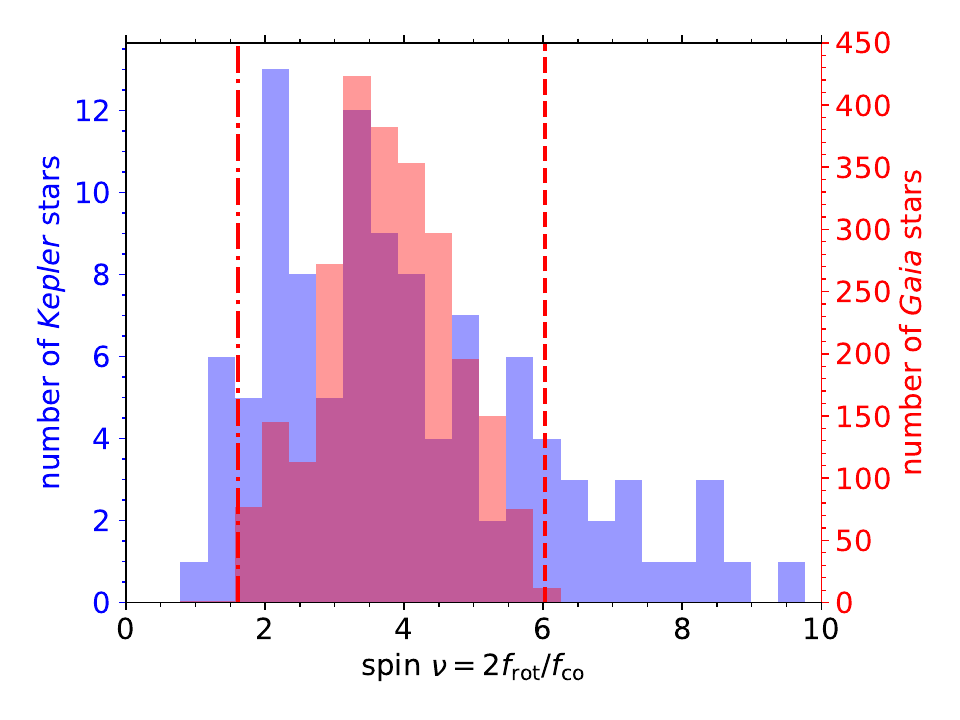}}}\\[-0.5cm]
\caption{Histograms of the dominant frequency of observed
  $(k,m)=(0,1)$ modes revealing an amplitude above 4\,mmag (top), of
  the near-core rotation frequency $f_{\rm rot}$ (middle), and of the
  spin parameter of the dominant mode (bottom) for the {\it Kepler\/}
  sample from \citet[][blue, left y-axis]{GangLi2020} and for our {\it
    Gaia\/} sample using Eq.\,(\protect{\ref{eq:frot-relation}}) (red,
  right y-axis).  The red dash-dotted and dashed lines mark the lower
  and upper boundaries for the {\it Gaia\/} sample, following from our
  selection criterion demanding that $f_{\rm
    dom}\in[0.7,3.2]\,$d$^{-1}$.
\label{fig:histograms}}
\end{figure}

A first result, represented by the histograms in
Fig.\,\ref{fig:histograms}, was that the distributions of the dominant
mode frequencies and of $f_{\rm rot}$ values are similar to those of
the much smaller {\it Kepler\/} sample of $\gamma\,$Dor stars in
\citet{GangLi2020} when we limit it to stars with dominant mode of
amplitude above 4\,mmag.  \citet{GangLi2020} derived the near-core
rotation from the slope of a period spacing pattern consisting of a
series of prograde dipole gravito-inertial modes of consecutive radial
order, following the method designed by \citet{VanReeth2016}. in this work, we
could  not rely on such period spacing patterns and we only made use of the
dominant mode in the blue ridge in Fig.\,\ref{fig:Dan} and of the recipe
in Eq.\,(\ref{eq:frot-relation}). As can be seen in
Fig.\,\ref{fig:histograms}, the distribution of $f_{\rm rot}$ achieved
in this simplified way for our large {\it Gaia\/} sample of
gravity-mode pulsators is fully in line with the one of the {\it
  Kepler\/} $\gamma\,$Dor pulsators.

Recall that we purposefully excluded the low-frequency regime in
the recipe, because the ultra-slow rotators among the gravity-mode
pulsators do not comply with the linear relationship in
Eq.\,(\ref{eq:frot-relation}). The dominant mode of all the 2,497 {\it
  Gaia\/} gravity-mode pulsators thus has a spin parameter placing
it into the sub-inertial regime. The values shown in the bottom
panel of Fig.\,\ref{fig:histograms} are in line with those of the 63
best characterised {\it Kepler\/} pulsators with spin parameters,
Rossby numbers, and stiffness values in \citet{Aerts2021-GIW}.

At the high frequency end, we excluded stars with $f_{\rm dom}$ above
3.2\,d$^{-1}$, reflected by the shorter tail in the distribution of
the spin parameter in the bottom panel of Fig.\,\ref{fig:histograms}.
We thus find by construction that none of the isolated gravity-mode
pulsators in our {\it Gaia\/} sample of field stars are expected to
rotate close to their critical rate, justifying the use of the TAR for
the recipe in Eq.\,(\ref{eq:frot-relation}).  We quantify this
qualitative statement in Sect.\,4. Here, we just note that at least
some of the $\gamma\,$Dor pulsators still embedded in their birth
cluster are known to rotate faster, as was found for the $\sim\!
100$\,Myr-old young open cluster NGC\,2516. \citet{GangLi2024}
identified 11 gravity-mode pulsators in this cluster and found most of
them to have near-core rotation frequencies $f_{\rm rot}\simeq
3$\,d$^{-1}$, which is faster than the rotation rates we find for the
population of isolated galactic {\it Gaia\/} field pulsators.

\begin{figure}
 \includegraphics[width=88mm]{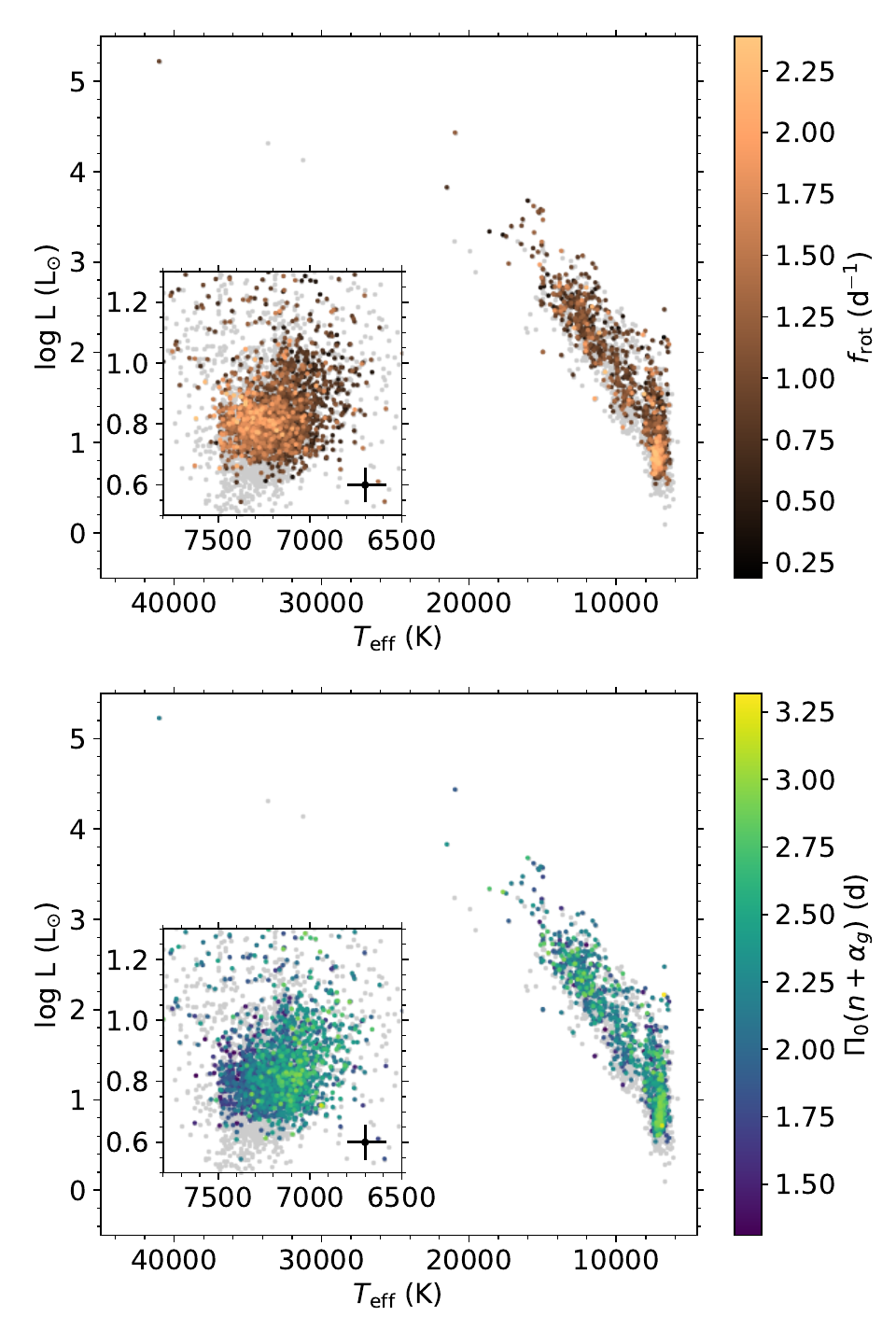}
 \caption{Our sample of \nstars {\em Gaia} DR3 gravity-mode pulsators
   in the HR diagram. The markers are coloured according to their $f_{\rm
     rot}$ (top) and $\Pi_0(n+\alpha_g)$ values (bottom). Other stars
   of the sample of 15,692 pulsators from \citet{MombargAerts2024} are
   shown by the grey markers. The inset axes show close-ups of the
   $\gamma$\,Dor instability region. Typical $3\sigma$ (average) error
   margins are shown by the black bars in the
   inset. \label{fig:Gaia_frot_HR}}
\end{figure}

In Fig.\,\ref{fig:Gaia_frot_HR} we place the stars in the
Hertzsprung-Russell (HR) diagram by colour-coding their markers
according to their $f_{\rm rot}$ and $\Pi_0\left(n+\alpha_g\right)$
values.  The pulsators cover a broad range in effective temperature
and luminosity, reflecting a wide range in stellar masses, which we
determine in Sect.\,4.  While the near-core rotation rates and scaled
buoyancy radii of the stars hotter than 8500\,K appear to be randomly
distributed, those of the stars within the $\gamma$\,Dor instability
region (shown in the inset) seem to correlate with $\log\,T_{\rm
  eff}$, in such a way that $f_{\rm rot}$ and $\Pi_0(n+\alpha_g)$
decrease and increase with $\log\,T_{\rm eff}$, respectively.

The 26 best modelled {\it Kepler\/} SPB stars did not reveal any
correlation between $\log\,T_{\rm eff}$ and the dominant oscillation
mode frequency, in line with earlier ground-based studies of 13 SPB
pulsators \citep[see Supplementary Material of][]{Pedersen2021}. We
confirmed this conclusion for the much larger {\it Gaia}
sample. This followed from simple linear regression analyses of the
form $y = a\log T_{\rm eff} + b$, where $y$ is taken to be $f_{\rm
  dom}$, $f_{\rm rot}$, and $\Pi_0\left(n+\alpha_g\right)$,
respectively. We also repeated the regression for the inverse quantities
and for the 10-base logarithm.  We dedicated Appendix\,A to illustrate
how these three seismic quantities relate to the effective temperature
for our {\it Gaia\/} sample. These relationships are graphically
depicted in the three scatter plots in
Fig.\,\ref{fig:Gaia_SPB_frot_Teff} for SPB stars and illustrate the
lack of any correlation for these pulsators.  The linear regressions
applied to all the stars above the $\gamma\,$Dor instability strip
have insignificant coefficients $(a,b)$ for all of $f_{\rm dom}$,
$f_{\rm rot}$, and $\Pi_0\left(n+\alpha_g\right)$.

\begin{table}[h!]
\tabcolsep=4pt
        \centering
    \caption{Regression coefficients for $y = a\log T_{\rm eff} + b$.}
        \label{tab:Gaia_frot_Teff}
        \begin{tabular}{lcccc} 
            \hline\hline\vspace{-9pt}\\
            $y$ & $a$ & $b$ & $r_{\rm Pearson}$ & $R^2$\\
            \hline\vspace{-9pt}\\\vspace{2pt}
            $f_{\rm dom}$ & $22.2_{-0.7}^{+0.7}$ & $-83.9_{-2.5}^{+2.6}$ & $0.426$ & 0.162\\\vspace{2pt}
            $f^{-1}_{\rm dom}$ & $-9.6_{-0.3}^{+0.3}$ & $37.74_{-1.11}^{+1.16}$ & $-0.426$ & 0.171\\\vspace{7pt}
            ${}^{10}\log\left(f_{\rm dom}\right)$ & $6.5_{-0.2}^{+0.2}$ & $-24.7_{-0.7}^{+0.7}$ & $0.426$ & 0.169\\\vspace{2pt}
            
            $f_{\rm rot}$ & $18.6_{-0.7}^{+0.7}$ & $-70.5_{-2.7}^{+2.7}$ & $0.400$ & 0.155 \\\vspace{2pt}
            $f^{-1}_{\rm rot}$ & $-18.5_{-0.8}^{+0.8}$ & $72.4_{-2.9}^{+3.1}$ & -$0.400$ & 0.166 \\\vspace{7pt}
            ${}^{10}\log\left(f_{\rm rot}\right)$ & $8.3_{-0.3}^{+0.3}$ & $-32.1_{-1.2}^{+1.2}$ & $0.400$ & 0.164 \\\vspace{2pt}
            
            $\Pi_0\left(n+\alpha_g\right)$ & $-16.3_{-1.5}^{+1.5}$ & $64.9_{-5.8}^{+6.0}$ & $-0.250$ & 0.187\\\vspace{2pt}
            $\left(\Pi_0\left(n+\alpha_g\right)\right)^{-1}$ & $3.9_{-0.4}^{+0.4}$ & $-14.5_{-1.4}^{+1.4}$ & $0.249$ & 0.186\\\vspace{2pt}
            ${}^{10}\log\left(\Pi_0\left(n+\alpha_g\right)\right)$ & $-3.5_{-0.3}^{+0.3}$ & $13.9_{-1.2}^{+1.4}$ & $-0.249$ & 0.188\\
            \hline
        \end{tabular}
\tablefoot{All listed coefficients $(a,b)$ differ significantly from zero at a
  $p$-value  of 5\%. Further, $r_{\rm Pearson}$ stands for 
  the Pearson correlation coefficient and $R^2$ is the coefficient
  of determination.}
\end{table}

As for the $\gamma\,$Dor mass regime, \citet[][their
  Fig.\,15]{GangLi2020} did not find any correlation between the
oscillation periods of the $(k,m)=(0,1)$ modes and $\log\,T_{\rm eff}$
for their sample of {\it Kepler\/} $\gamma\,$Dor stars. No clear
correspondence between $\Pi_0$ and $\log\,T_{\rm eff}$ was found
either. We did find weak correlations from our much larger {\it Gaia\/}
sample but the comparisons are not equivalent. Indeed,
\citet{GangLi2020} were able to identify the radial orders of the
modes from detected period spacing patterns for each of the
pulsators. For our work, we did not have any information on the radial order
$n$ of the dominant mode, so we could not make use of the buoyancy radius
$\Pi_0$ itself. Rather, we had to use a scaled value including the
unknown factor $(n+\alpha_g)$. Because we expected weak correlations
between the seismic quantities and $\log\,T_{\rm eff}$ at best, we
computed Pearson's correlation coefficients and conducted linear
regressions with Monte Carlo sampling analyses for 10,000 iterations
to assess the uncertainties. We selected the stars with $T_{\rm eff}$
values between 6810\,K and 7540\,K following the local target density
in the HR diagram in Fig.\,\ref{fig:Gaia_frot_HR} to ensure there to
be at least 3 sample stars per 10\,K within this $T_{\rm eff}$
range. In each iteration, we sampled new values for $\log\,T_{\rm
  eff}$, $f_{\rm dom}$, $f_{\rm rot}$, and $\Pi_0(n+\alpha_g)$ for
85\,\% of the selected targets and fitted linear, inverse, and
logarithmic relations using Theil-Sen estimators
\citep[e.g. ][]{Theil1950,Sen1968}. The results are summarised in
Table\,\ref{tab:Gaia_frot_Teff} and visualised in
Fig.\,\ref{fig:Gaia_frot_Teff}, where we point out that the dearth of
stars in the upper panel at $f_{{\rm dom,}\ \geq\,4\,{\rm
    mmag}}\simeq\,1\,$d$^{-1}$ is due to the spectral window of the
TESS data as illustrated by \citet{HeyAerts2024}.  Although the $R^2$
values are low, as anticipated from the earlier {\it Kepler\/} results
by \citet{GangLi2020}, we find that $f_{\rm dom}$, $f_{\rm rot}$, and
$\Pi_0(n+\alpha_g)$ all correlate modestly but significantly with
$\log\,T_{\rm eff}$.  The positive $a$ values in
Table\,\ref{tab:Gaia_frot_Teff} point to a positive correlation
between $f_{\rm dom}$ and $\log\,T_{\rm eff}$, and by implication also
between $f_{\rm rot}$ and $\log\,T_{\rm eff}$. These are well visible
in Fig.\,\ref{fig:Gaia_frot_Teff}. On the other hand, the negative $a$
value for $\Pi_0(n+\alpha_g)$ points to an anti-correlation with
$\log\,T_{\rm eff}$ (see bottom panel of
Fig.\,\ref{fig:Gaia_frot_Teff}). The weaker correlation occurs because
of the wider asymmetrical confidence regions for this quantity as
revealed in Fig.\,\ref{fig:Kepler-relation-Pi0n}.

While the gravity-mode pulsators above the $\gamma$\,Dor instability
region cover a wide range of stellar masses \citep[][see also
  Sect.\,4]{MombargAerts2024}, those inside the narrow instability
region have similar mass. This allowed us to probe the evolution of
these stars more easily \citep[e.g.,][]{Mombarg2021,Fritzewski2024b}
than the one of the less populated SPB sample.  The modest yet
significant linear decrease of $f_{\rm rot}$ as $\log\,T_{\rm eff}$
decreases for the $\gamma\,$Dor pulsators offers observational input
to evaluate angular momentum transport theories and simulations as
these stars evolve along the main sequence
\citep[e.g. ][]{Ouazzani2019,Aerts2019-araa,Aerts2021}.  Even if we
could compute more complex statistical models than the simple linear
ones used here, we refrained from doing so as long as we do not have the
outcome of forward modelling based on fitting individual identified
modes from grids of models with a variety of input physics as done for
SPB stars by \citet{Pedersen2021}. Indeed, \citet[][their
  Fig.\,16]{GangLi2020} have illustrated that the connection between
$f_{\rm rot}$, $\Pi_0$ and $T_{\rm eff}$ is complex and depends on the
details of transport processes in the stellar interior as the stars
evolve. We plan to achieve more advanced asteroseismic and statistical
modelling in future work, by fully exploiting a considerable number of
the identified modes from the TESS light curves of suitable
$\gamma\,$Dor pulsators, rather than working with only the dominant
mode as done here for the large {\it Gaia\/} sample. The results in
Table\,\ref{tab:Gaia_frot_Teff} are promising to undertake such
time-consuming modelling in the future.

Finally, even if the correlation is weak, we found that the scaled
buoyancy radius $\Pi_0(n+\alpha_g)$ increases with decreasing
$\log\,T_{\rm eff}$. This is contrary to $\Pi_0$ itself, which
decreases as the star cools
\citep[e.g. ][]{Mombarg2019,Ouazzani2019}. We thus found observational
evidence of an increase in the radial order of the dominant excited
mode as the $\gamma$ Dor stars evolve.

\section{Grid-based forward modelling}
\begin{figure*}[h!]
\sidecaption
    \includegraphics[width=12cm]{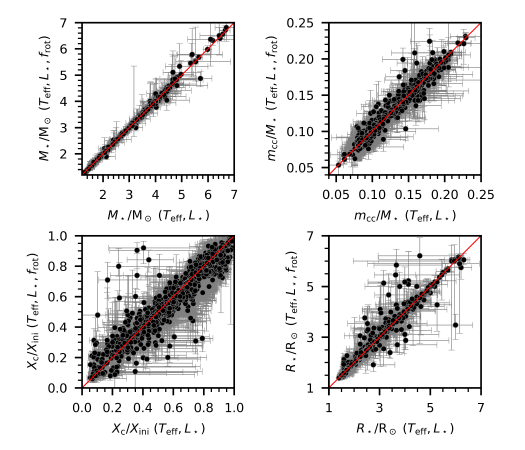}
    \caption{Derived measurements for the stellar mass (top left), the
      fractional convective core mass (top right), the core
      hydrogen-mass fraction compared to the initial hydrogen-mass
      fraction (bottom left), and the stellar radius (bottom right) of
      the 2,464 gravity-mode pulsators modelled via the grid with solar
      metallicity. The quantities on the abscissa are derived using
      the effective temperature and luminosity as input, while for the
      quantities on the ordinate, the near-core rotation frequency is
      added as an extra observable. The red lines indicate perfect
      correspondence.}
    \label{fig:modelling}
\end{figure*}

\cite{Mombarg2024} and \citet{Fritzewski2024b} illustrated how
measured values of the near-core rotation frequency and $\Pi_0$
provide useful extra constraints, in addition to the luminosity and
effective temperature of a star, for stellar modelling based on
evolutionary tracks. In these two studies, the modelling had the
purpose of determining the stellar mass $M_\star$, the core
overshooting leading to $m_{\rm cc}$, and the age via its proxy $X_{\rm
  c}/X_{\rm ini}$. Such evolutionary modelling is a lot less
cumbersome (but also less powerful) than detailed asteroseismic
modelling of individual mode frequencies
\citep[e.g. ][]{Moravveji2015,Moravveji2016,Mombarg2020,Mombarg2021,Pedersen2021}.
This simplified evolutionary modelling by \cite{Mombarg2024} and
\citet{Fritzewski2024b} was done by relying on grids of stellar
evolution models calibrated to $\gamma$~Dor star asteroseismology
computed by \citet{Mombarg2023-AM}.

In recent applications of gravity-mode asteroseismic modelling, the
near-core rotation frequency of the star was derived from 4-years long
{\it Kepler\/} light curves. These typically lead to much better
precision for $f_{\rm rot}$ compared to the precision of $\sim 20\%$
we achieved here from applying Eq.\,(\ref{eq:frot-relation}) to the
measured $f_{\rm dom}$, after identification of the mode as
$(k,m)=(0,1)$ in the sub-inertial regime. For the current study,
we investigated whether
the addition of an estimated near-core rotation frequency from just
the dominant pulsation frequency helps constrain the fundamental
stellar parameters $M_\star$, $m_{\rm cc}$, $R_\star$, and $X_{\rm
  c}/X_{\rm ini}$.

We applied the same methodology as presented in
\citet{MombargAerts2024}, where a conditional neural flow trained on a
grid of rotating stellar evolution models computed with \texttt{MESA}
\citep{Paxton2011, Paxton2013, Paxton2015, Paxton2018, Paxton2019,
  Jermyn2023} was used as an interpolator. We refer to
\citet{MombargAerts2024} for the details on how this methodology
works. For the sub-sample of stars with a measured $f_{\rm rot}$
obtained in this work, we re-derived their $M_\star$, $m_{\rm cc}$,
$R_\star$, and $X_{\rm c}/X_{\rm ini}$ by adding $f_{\rm rot}$ as a
third observational constraint, in addition to the two observables
$\log T_{\rm eff}$ and $\log(L/L_\odot)$ used for the modelling by
\citet{MombargAerts2024}.  As in that paper, we relied on two extensive
grids of stellar models, for values of the metallicity equal to $Z =
0.0045$ ($[{\rm M/H}] = -0.5$) and $Z = 0.014$ ($[{\rm M/H}] =
0.0$). These two metallicities were chosen because we do not have a
homogeneously derived measurement of the metallicity for all the
sample stars. The two grids encapsulate the measured range of $Z$ for
gravity-mode pulsators having a metallicity estimate from
high-resolution ground-based spectroscopy \citep{Gebruers2021} and
from {\it Gaia\/} spectroscopy \citep{deLaverny2024}.  In addition to
these two grids from \citet{MombargAerts2024}, we computed a third
grid for $Z=0.008$ ($[{\rm M/H}] = -0.25$) and trained a conditional
neural flow on this grid, following the same methodology as described
in that paper. From the perspective of position in the HR diagram,
2,375 pulsators were covered by the lowest-metallicity grid, 2,439 by the
intermediate-metallicity grid and 2,464 by the solar metallicity grid.
We worked with these stars from here onwards.

Figure~\ref{fig:modelling} shows the difference in the derived
fundamental stellar quantities depending on whether $f_{\rm rot}$ is
included as an additional constraint or not, for the 2,464 pulsators
captured by the solar metallicity grid. As can be seen in the top-left
panel, the derived stellar mass is consistent between the two
cases. The largest scatter is observed in the evolutionary stage
($X_{\rm c}/X_{\rm ini}$, bottom-left panel), which is harder to
constrain than the mass as it depends strongly on (unknown) element
mixing in the models. In the context of ensemble modelling of
gravity-mode pulsators, we found that $f_{\rm rot}$ derived from
Eq.\,(\ref{eq:frot-relation}) applied to the {\it Gaia\/} light curves
helps to refine the determination of the evolutionary stage of the
star from rotating stellar models. Furthermore, with the inclusion of
$f_{\rm rot}$, we achieved slightly more precise estimates of $X_{\rm
  c}/X_{\rm ini}$, going from average (absolute) uncertainties of
0.10-0.08 to 0.08-0.06. In case of the convective core mass (top-right
panel) and radius (bottom-right panel), the majority of the stars with
the largest difference still have measurements consistent within the
68\% confidence interval. The average uncertainties $m_{\rm
  cc}/M_\star$ and $\log(R_\star/R_\odot)$ remain approximately the
same when $f_{\rm rot}$ is included in the modelling or not. This
re-affirmed that estimation of $f_{\rm rot}$ can be decoupled from
forward asteroseismic modelling as emphasised by \citet{VanReeth2016},
\citet{Ouazzani2017} and \citet{Aerts2018}.

The comparative results for the 2,375 pulsators covered by the
lower-metallicity grids were similar, keeping in mind that the mass,
luminosity, and metallicity are tightly connected for main-sequence
stars \citep[see their strong relationships deduced
  by][]{MombargAerts2024}. We omit them here for brevity, but they are
shown visually in Figs\,\ref{fig:modelling-metalpoor} and
\ref{fig:modelling-m025} in Appendix\,A. Given that we do not have
measurements of $Z$ for the sample stars, we provide the stellar
parameters for the three grids in electronic format at the CDS,
following the format in Tables\,\ref{table-solar}, \ref{table-Z0-008},
and \ref{table-Z0-0045} (see also footnote to the title of the paper).

The differences in the stellar parameters deduced from the grid
modelling for the stars treated by the three grids reflect the
systematic uncertainties for the (core) masses, radii, and
evolutionary stages induced solely by the differences of 0.0060 and
0.0095 in $Z$ ({-0.25} and {-0.50} in [M/H]).  Again, these
differences are systematic and follow the tight (core) mass --
luminosity -- radius -- metallicity relationships. They range up to
0.55\,M$_\odot$ in mass, up to 0.28\,M$_\odot$ in convective core
mass, and from about {$-0.25$}\,R$_\odot$ to {$+0.25$}\,R$_\odot$ in
radius. By construction, the evolutionary stages differ from each
other because the grid having $Z=0.0045$ was built up by
\citet{MombargAerts2024} to make sure all observed stars are above the
ZAMS and to encapsulate the range in unknown metallicity reported in
the literature for gravity-mode pulsators, affecting the evolution
appreciably.  For the grid with $Z=0.008$ the differences in $X_{\rm
  c}/X_{\rm ini}$ remain more modest for most stars, reaching up to
0.3 compared to the solar metallicity case.  A visual representation
of the distributions of these systematic differences for all the stars
with overlapping grid solutions is shown in
Fig.\,\ref{fig:syshistograms} and reflects the well-known
luminosity--mass-radius--metallicity relationships for main-sequence
stars \citep[cf.\,][]{MombargAerts2024}.

\begin{figure*}[h!]
\sidecaption
    \includegraphics[width=12cm]{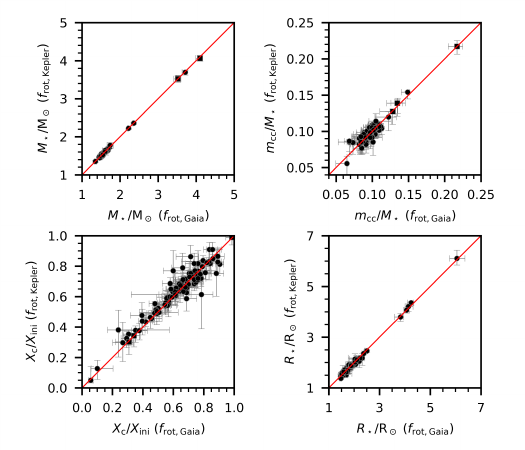}
    \caption{Derived measurements for the stellar mass (top left), the
      fractional convective core mass (top right), the core
      hydrogen-mass fraction compared to the initial hydrogen-mass
      fraction (bottom left), and the stellar radius (bottom right) of
      96 $\gamma\,$Dor stars (circles) and three SPB stars (squares)
      modelled via the grid with solar metallicity. The quantities are
      derived using the effective temperature, luminosity, and
      near-core rotation frequency from the dominant dipole prograde
      mode detected in the {\it Gaia\/} DR3 light curves (x-axis)
      versus the higher-precision value deduced from the slope of a
      period spacing pattern of dipole prograde modes of consecutive
      radial order following \citet{GangLi2020} and
      \citet{Pedersen2021} (y-axis).}
    \label{fig:KeplerGaiafrot}
\end{figure*}

Finally, we also studied the effect of having the higher-precision input
of $f_{\rm rot}$ from the slope of a period-spacing pattern of dipole
prograde modes based on 4-year {\it Kepler\/} light curves instead of
the value based on just $f_{\rm dom}$ from {\it Gaia\/} DR3 light
curves. We considered the 105 {\it Kepler\/} pulsators from
\citet{GangLi2020} and \citet{Pedersen2021} used to construct
Eq.\,(\ref{eq:frot-relation}) and shown in
Fig.\,\ref{fig:Kepler-relation-frot}. We computed their luminosity and
effective temperature in the same way as we did it for the {\it
  Gaia\/} sample and found 96 of the $\gamma\,$Dor stars with dominant
$(k,m)=(0,1)$ mode to be covered by the model grid with solar
metallicity, while 3 of the SPB stars. We show the results of the
grid modelling based on the less precise {\it Gaia\/} $f_{\rm rot}$
and the more precise {\it Kepler\/} $f_{\rm rot}$ estimates for these
99 {\it Kepler\/} pulsators in Fig.\,\ref{fig:KeplerGaiafrot}. The
four estimated parameters have very consistent values for almost all
99 pulsators.

Having deduced the stellar parameters from grid modelling, we are now
able to come back to one of the main assumptions in our procedure,
namely that the pulsators may rotate fast, yet such that the
centrifugal force can still be neglected. The centrifugal force
affects both the equilibrium models and the pulsation
computations. The former become oblate spheroids rather than spheres,
while the modes will not only get affected by the Coriolis force but
as well by the centrifugal force $\sim\Omega^2$. To evaluate the
impact on the equilibrium models, we tested at which fraction of the
critical rotation rate the stars are rotating. We note that different
definitions for the critical rotation rate are used in the literature,
notably the Kepler and Roche critical rate. In practical applications,
these differ by a factor up to 1.84 due to the use of the equatorial
or polar radius of the star at the critical rate \citep[see][for
  discussions]{Rieutord2016,AertsTkachenko2024}. Here, we worked with
the cyclic Kepler critical rotation rate defined as $f_{\rm
  crit}\equiv\sqrt{GM_\star/(R_\star^{\rm eq})^3}$ and assumed the
derived radii of the stars to correspond with their equatorial
value. This is an appropriate choice since we are dealing with stars
having a dominant sectoral dipole mode in an observer's frame, which
favours a more equator-on than pole-on view. We then used the mass
result from the grid modelling to compute $f_{\rm rot}/f_{\rm crit}$
for all the pulsators in our {\it Gaia\/} sample. The values ranged
from 0.7\% to 25\% and imply that the stars are not strongly deformed
by the rotation according to the asteroseismic validity criteria for
prograde modes adopted by \citet{Ballot2012} and
\citet{Ouazzani2017}. The latter authors, as well as
\citet{Ballot2013} concluded that the centrifugal force can safely be
ignored for the calculation of prograde dipole gravity modes when
$f_{\rm rot}/f_{\rm crit}<40\%$, which is fulfilled for all our sample
stars.  This justifies the use of the TAR also from the perspective of
the stellar deformation due to the rotation.

Overall, our results reveal the great potential of using
Eq.\,(\ref{eq:frot-relation}) to deduce $f_{\rm rot}$ and grid-based
stellar parameter estimation based on just the value of the dominant
mode frequency from {\it Gaia\/} light curves along with the effective
temperature and luminosity.  All these quantities are homogeneously
computed and available from {\it Gaia\/} DR3 (and its future Data
Releases DR4 and DR5). When coupled to an asteroseismically calibrated
grid of stellar models as those in \citet{MombargAerts2024} and used
here, these observables deliver stellar parameters calibrated by the
star's internal properties via the observed dominant mode frequency.
The only requirement for the procedure to work is the ability to
identify pulsators with a dominant dipole prograde mode having its
frequency in the frame co-rotating with the star, $f_{\rm co}$, in the
sub-inertial regime of the frequency spectrum (spin parameter
$\nu=2\cdot f_{\rm rot}/f_{\rm co}>1$), while ensuring a modest
rotational deformation of $f_{\rm rot}/f_{\rm crit}<40\%$.

\section{Near-core to surface rotation estimates}

\subsection{Evolution of the near-core rotation}
Combining the results of the previous two sections, we studied
the near-core rotation frequency as a function of evolutionary
stage. Asteroseismology has demonstrated that angular momentum
transport from the deep interior to the envelope of stars happens more
efficiently than anticipated from theories of transport processes
adopted prior to the era of space asteroseismology
\citep{Ouazzani2019}. It was also found that gravity-mode pulsators
have low levels of near-core to surface rotation \citep[][for
  observational
  summaries]{Aerts2019-araa,GangLi2020,Aerts2021,AertsTkachenko2024}. Moreover,
stars born with a radiative envelope lose angular momentum rapidly
beyond the TAMS (at $X_c/X_{\rm ini}=0)$), while their near-core
rotation drops appreciably \citep[][Fig.\,6]{Aerts2021}.

\begin{figure}
    \centering
    \includegraphics[width=0.42\textwidth]{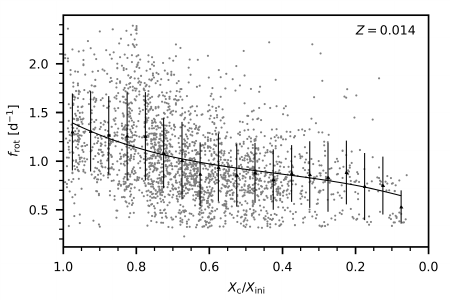}
    \includegraphics[width=0.42\textwidth]{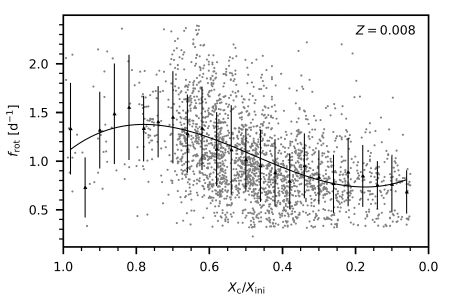}
    \includegraphics[width=0.42\textwidth]{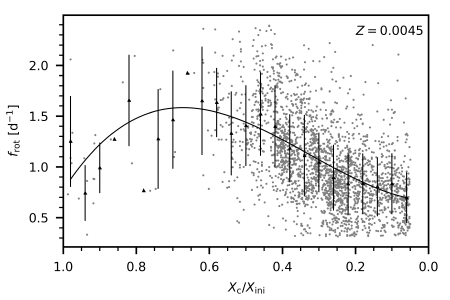}
    \includegraphics[width=0.42\textwidth]{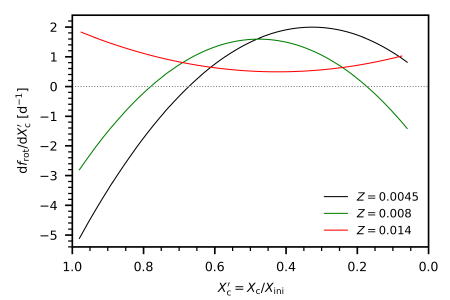} \caption{
      {Inferred near-core rotation frequencies of the {\it Gaia\/}
        gravity-mode pulsators plotted as a function of evolutionary
        stage (grey dots), assuming solar metallicity (2,464 stars, top
        panel), ${\rm [M/H]}=-0.25$ (2,439 stars, second panel), and
        ${\rm [M/H]}=-0.5$ (2,375 stars, third panel). Solid lines
        indicate spline fits through binned averages (see text). The
        bottom panel shows the derivatives of the splines from the
        three upper panels, indicating the change in the near-core
        rotation frequency with respect to the fraction of hydrogen in
        the core. The dotted grey line separates positive and negative
        values.}}
    \label{fig:frotstage}
\end{figure}

In Fig.\,\ref{fig:frotstage}, we show the model-independent near-core
rotation frequency, $f_{\rm rot}$, as a function of the
model-dependent evolutionary stage for our sample of gravity-mode
pulsators, for the three grids.  Thanks to the four times larger size of
our {\it Gaia\/} population of pulsators with a measurement of $f_{\rm
  rot}$ compared to the \citet{GangLi2020} {\it Kepler\/} sample, and
to our estimate of the evolutionary stage, we see a clear decline of
the average near-core rotation frequency of the population as the
stars evolve.

We fitted uni-variate splines through bin-averaged $f_{\rm rot}$ values,
where we selected the optimal bin size in $X_{\rm c}^\prime = X_{\rm
  c}/X_{\rm ini}$ according to Scott's rule \citep{Scott1979}. This
rule states that a histogram with bin size proportional to
$3.49\cdot\sigma/(N_{\rm points}^{1/3})$, where $N_{\rm points}$ is
the number of points in the bins and $\sigma$ the standard deviation
assuming a normal distribution, is very efficient in approaching the
true distribution.  The smoothing factor for the splines was chosen
such that higher values do no longer change the resulting derivative
${\rm d}f_{\rm rot}/{{\rm d}X_{\rm c}^\prime}$. The derivatives of the
fitted splines are shown in the bottom panel of
Fig.~\ref{fig:frotstage}.  When assuming $Z = 0.014$, we found purely
decreasing $f_{\rm rot}$ values during the main sequence.  Despite the
large range of near-core rotation frequencies per evolutionary stage
-- reflecting the variety of initial conditions at birth for this
population of field stars -- the average near-core rotation per bin in
$X_c/X_{\rm ini}$ shows a gradual decline from the zero-age to the
terminal-age main sequence, with roughly a factor of two.

Assuming $Z = 0.0045$, the results were less clear. We found that stars
on average have an increasing $f_{\rm rot}$ (negative derivative)
during the first part of the main sequence, followed by a decrease in
$f_{\rm rot}$. Yet, we note that the part where ${\rm d}f_{\rm
  rot}/{{\rm d}X_{\rm c}^\prime} < 0$ for $Z = 0.0045$ is severely
undersampled and so we could not rely on these results. However, the same
decline of $f_{\rm rot}$ with a factor two as for the solar
metallicity grid was obtained for the regime $X_{\rm c}/X_{\rm
  ini}<0.5$ when the stars are modelled with the metal-poor grid.
{When assuming $Z=0.008$, a similar trend in ${\rm d}f_{\rm rot}/{{\rm
      d}X_{\rm c}^\prime}$ is observed.}

Finally, we checked that the trends found become less clear but do not
change globally when we split up the observed sample in stars with an
initially growing convective core and stars whose convective core
shrinks as of the ZAMS. The distinction between these two cases
happens for masses between about 1.6\,M$_\odot$ and 1.8\,M$_\odot$,
depending on the metallicity. We therefore conclude to have found
observational evidence that the near-core rotation frequency of
isolated gravity-mode field pulsators in the Milky Way decreases with
a factor of about two by the time they will reach the TAMS if we
consider all the well covered bins in Fig.~\ref{fig:frotstage},
irrespective of the initial metallicity.

\subsection{Limits on the surface rotation}

Only 58 of the {\it Kepler\/} $\gamma\,$Dor stars also have a value
for the surface rotation, measured from rotational modulation
\citep{VanReeth2018,GangLi2020} and only 22 have a measurement of
their averaged envelope rotation, $f_{\rm env}$, determined from
identified rotational splitting
\citep{Kurtz2014,Saio2015,Keen2015,GangLi2019}. For the SPB stars,
information is even more restricted, with only one star having clear
rotational splitting \citep{Papics2014} and a rotational profile
throughout the star from frequency inversions \citep[the
  3.3\,M$_\odot$ SPB KIC\,10526294,][]{Triana2015}. These limited
observational constraints on differential near-core to surface
rotation in main-sequence stars of intermediate and high mass
\citep[as summarised by][Fig.\,6]{Aerts2021} restrict our ability to
improve angular momentum transport theories, as these rely on the
gradient of the rotation profile throughout the star
\citep{Ouazzani2019,Aerts2019-araa}.

We were unable to assess the envelope or surface rotation from the {\it
  Gaia\/} light curves. However, we derived a lower limit for the
cyclic surface rotation frequency, $f_{\rm surf}$, from the radius
estimates shown in Fig.\,\ref{fig:modelling} for those stars with a
measurement of the spectral line broadening offered by the {\it
  Gaia\/} spectroscopy \citep{Fremat2023}. \citet{Aerts2023} showed
the {\it Gaia\/} {\tt vbroad} parameter to capture the joint effect of
time-independent rotational line broadening and time-dependent
tangential pulsational broadening due to gravity modes. Moreover,
pulsational line broadening of gravity-mode pulsators is typically
only a small fracion of the rotational broadening for moderate to fast
rotators \citep{DeCatAerts2002,Aerts2004,DeCat2006}. Hence {\it
  Gaia}'s {\tt vbroad} measurement is a meaningful approximation of
$2\pi R_\star\cdot f_{\rm surf}\cdot\sin i$ for such stars, where $i$
is the (unknown) inclination angle of the rotation axis of the star in
the line-of-sight.

A total of 969 of our sample stars have a significant {\tt vbroad}
measurement, by which we mean it is above zero at 1-$\sigma$ level,
and a radius estimate from the solar metallicity grid. For the
lower-metallicity grids it concerns { 965 stars for $Z=0.008$ and 929
  stars for $Z=0.0045$.}  Using the radius estimates from the grid
modelling, we computed $f_{\rm surf}\cdot\sin i$ from {\tt vbroad}. As
this is a lower limit for the true surface rotation frequency $f_{\rm
  surf}$, we obtained an upper limit for the near-core to surface
differential rotation from calculating $f_{\rm rot}/(f_{\rm
  surf}\cdot\sin i)$. Our results are illustrated in
Fig.\,\ref{fig:upperlimits} and reveal values up to 5.4, irrespective
of the assumption about the metallicity.  The upper limits shown in
Fig.\,\ref{fig:upperlimits} have large uncertainty (up to 100\%) for
many of the stars due to the large errors of {\tt vbroad} (not shown
in the figure for visibility reasons).  So far, asteroseismology of
single $\gamma\,$Dor stars covering the mass range
$[1.3,1.9]\,$M$_\odot$ has shown them all to have quasi-rigid rotation
at the level of $f_{\rm rot}/f_{\rm surf}\in [0.9,1.1]$ if such a
measurement is available \citep[][for a
  summary]{GangLi2020,Aerts2021}.  Our work increases the mass range
of stars with an estimate of differential rotation, even if we can
provide only an upper limit. Indeed, of the 969 gravity-mode pulsators
presented in Fig.\,\ref{fig:upperlimits}, 58 have a mass between
2.0\,M$_\odot$ and 4.4\,M$_\odot$ when adopting solar metallicity.
The unknown factor $\sin i$ and the large uncertainty of {\tt
  vbroad} did not allow us to deduce stricter constraints than just a
rough upper limit for $f_{\rm rot}/f_{\rm surf}$. Measurements of the
actual ratio $f_{\rm rot}/f_{\rm env}$ can be investigated from
detailed analyses of the TESS light curves for the hybrid low-order
pressure and high-order gravity mode pulsators in our sample. This
will be taken up in future work.

\section{Conclusions and outlook}

In this work, we derived the internal rotation frequency of 2,497
gravity-mode pulsators discovered recently from {\it Gaia\/} DR3 light
curves. These field stars cover the mass range from 1.3\,M$_\odot$ to
about 7\,M$_\odot$ and the entire main sequence. We provide an
easy-to-apply procedure to deduce their near-core rotation frequency.
Our linear regression recipe is based on the dominant prograde dipole
oscillation mode found in the {\it Gaia\/} DR3 light curves.  The
2,497 {\it Gaia\/} pulsators have near-core rotation rates ranging
from 0.3\,d$^{-1}$ ($\simeq\! 3.5\,\mu$Hz) to 2.4\,d$^{-1}$ ($\simeq\!
27.8\,\mu$Hz), with a distribution in line with those of {\it
  Kepler\/} gravity-mode pulsators. All these {\it Gaia\/} pulsators
have an identified dominant dipole prograde mode in the sub-inertial
frequency regime with spin parameters ranging from about 2 to 6.  They
cover a ratio of the near-core rotation rate to the Keplerian critical
rate from 0.7\% to 25\%.  Their stellar parameters were deduced from
asteroseismically calibrated grids of rotating stellar models.  With
our work, we quadruple the sample size of intermediate-mass dwarfs
with a measurement of the near-core rotation frequency in the
transition layer between the convective core and the radiative
envelope.

The weakest point of the grid modelling performed for the {\it Gaia\/}
sample is the quite large uncertainty for the evolutionary stage of
the stars, quantified as the central hydrogen mass fraction over the
initial value, $X_c/X_{\rm ini}$. The reason is that we do not have a
good estimate of the initial metallicity of the pulsators, preventing
us to assign an age to the stars based on their $X_c/X_{\rm ini}$
value. This can perhaps be overcome from future detailed asteroseismic
modelling based on a measurement of abundances from high-resolution
spectroscopy coupled to the fitting of numerous identified oscillation
mode frequencies from grids of stellar models built with a variety of
choices for the input physics. The study by \citet{Pedersen2021} is so
far the only one that achieved such an asteroseismic age calibration
for gravity-mode pulsators on the main sequence. The authors
considered eight grids of models with different input physics for the
transport processes, but their study only covered 26 modelled SPB
stars. Future similar work for a large sample with the most promising
among our {\it Gaia\/} pulsators from their complete list of
significant TESS oscillation frequencies is on the horizon.

For 969 pulsators in the sample, we derived an upper limit of the
radial differential rotation between the boundary of the convective
core and the surface, by relying on the {\it Gaia\/} DR3 {\tt vbroad}
observable combined with the radius estimate from the grid
modelling. We find values up to 5.4.  These results cover the mass
range of $[1.3,4.4]\,$M$_\odot$ and constitute an interesting
sub-sample to calibrate stellar evolution theory and angular momentum
transport theories for rotating main-sequence stars with a convective
core and a radiative envelope.  This is of particular interest during
the initial $\sim 20\%$ of the main sequence. In order to understand
such early phases of stellar evolution, it is needed to deduce
high-precision values for $X_c/X_{\rm ini}$. An excellent way to
achieve this, as well as calibrate age-dating from $X_c/X_{\rm ini}$,
is to perform asteroseismic modelling of gravity-mode pulsators in
detached binaries and/or open clusters. Such modelling work was done
for a few binaries \citep{Schmid2016,Sekaran2021,Kemp2024} and also
for the open cluster UBC\,1 \citep{Fritzewski2024a}. It is currently
ongoing for the very young ($\simeq\! 100\,$Myr) open cluster
NGC\,2516, in which \citet{GangLi2024} discovered 9 $\gamma\,$Dor and
2 SPB pulsators rotating at half their critical Keplerian rotation
rate. This is a higher internal rotation regime than the one covered
by our sample of {\it Gaia\/} field pulsators.

Finally, we point to the large potential of future {\it Gaia\/} data
releases. Combining these data with high-cadence high-precision space
photometry from the ongoing TESS and future PLATO \citep{Rauer2024}
missions will allow us to scale up the work presented here to millions
of pulsators, instead of thousands.  Our recipe in
Eq.\,(\ref{eq:frot-relation}) to deduce the near-core rotation of
pulsators with a dominant prograde dipole gravito-inertial mode having
a mass above 1.3\,M$_\odot$ is suitable to facilitate fast
asteroseismic grid modelling. At the same time, it is also a
particularly handy and relevant tool for exoplanet host studies in
terms of the angular momentum properties of the hottest among the
main-sequence star-planet systems.


\begin{acknowledgements}
We thank the referee for the useful detailed suggestions, which helped
us to improve the presentation of our work. CA is grateful to Maarten
Dirickx, Toon De Prins and Thomas Ceulemans for helping her with
compiler issues.  The research leading to these results has received
funding from the { KU\,Leuven Research Council (grant C16/18/005:
  PARADISE) and from the } European Research Council (ERC) under the
Horizon Europe programme (Synergy Grant agreement N$^\circ$101071505:
4D-STAR).  While partially funded by the European Union, views and
opinions expressed are however those of the authors only and do not
necessarily reflect those of the European Union or the European
Research Council. Neither the European Union nor the granting
authority can be held responsible for them.  CA also acknowledges the
Belgian Federal Science Policy Office (BELSPO) for their financial
support in the framework of the PRODEX Programme of the European Space
Agency (ESA), facilitating the exploitation of the {\it Gaia\/} data.
JSGM acknowledges funding fron the French Agence Nationale de la
Recherche (ANR), under grant MASSIF (ANR-21-CE31-0018-02).  The
authors appreciated valuable comments from Dominic Bowman, Dario
Fritzewski, and Mathijs Vanrespaille on an early version of the
manuscript.
\end{acknowledgements}

\bibliographystyle{aa}
\bibliography{20250220-RMRA.bib}

\appendix

\section{Correlations between seismic quantities and the effective temperature}
\begin{figure}[h!]
  \centering
 \includegraphics[width=6.cm]{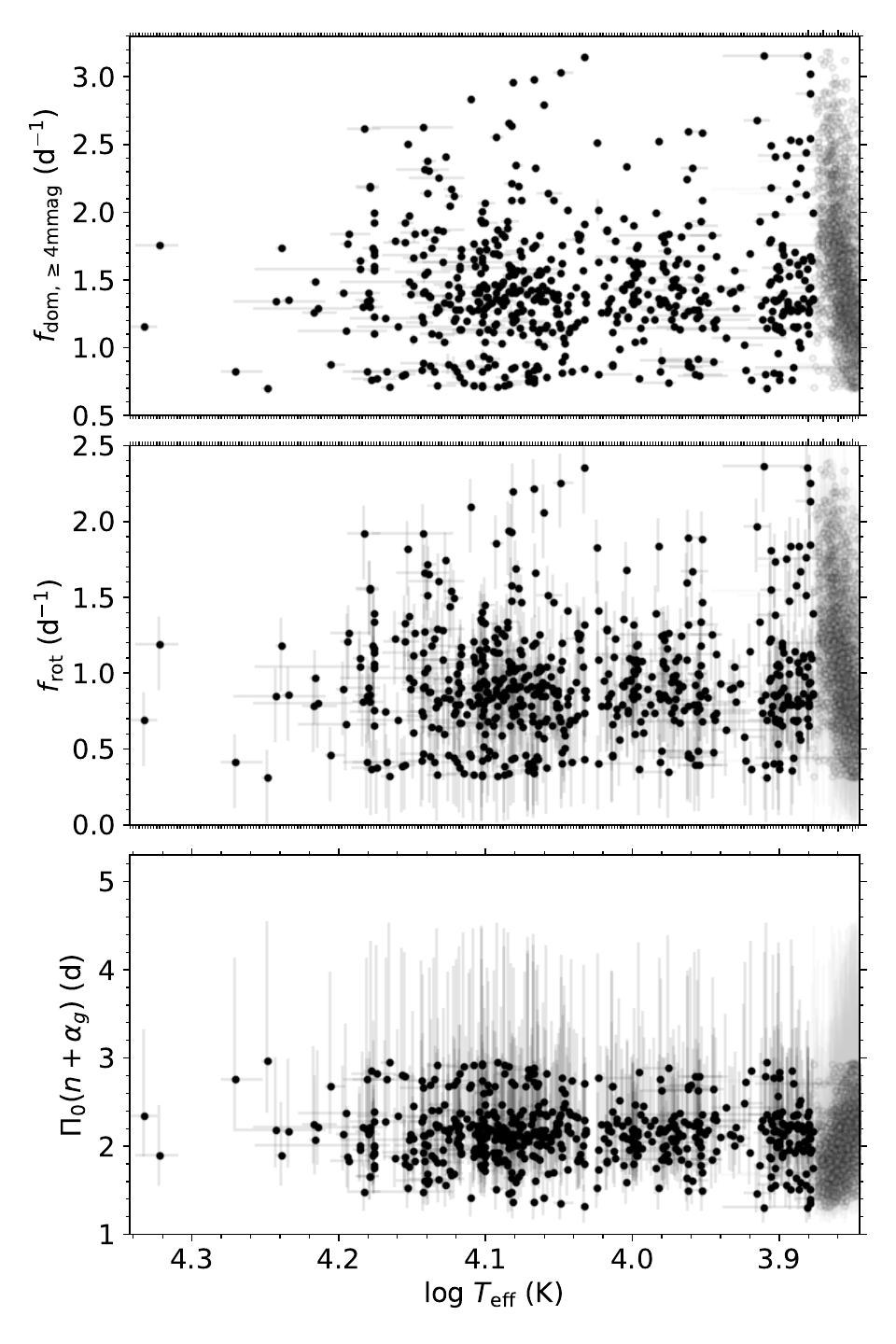}
 \caption{Dominant mode frequency (top), inferred near-core rotation
   frequency $f_{\rm rot}$ (middle), and
   $\Pi_0\left(n+\alpha_g\right)$ (bottom) plotted as a function of
   $\log\,T_{\rm eff}$ for pulsators above (black) and within (grey)
   the $\gamma$\,Dor instability region.
\label{fig:Gaia_SPB_frot_Teff}}
\end{figure}
\begin{figure}[h!]
  \centering
 \includegraphics[width=6.cm]{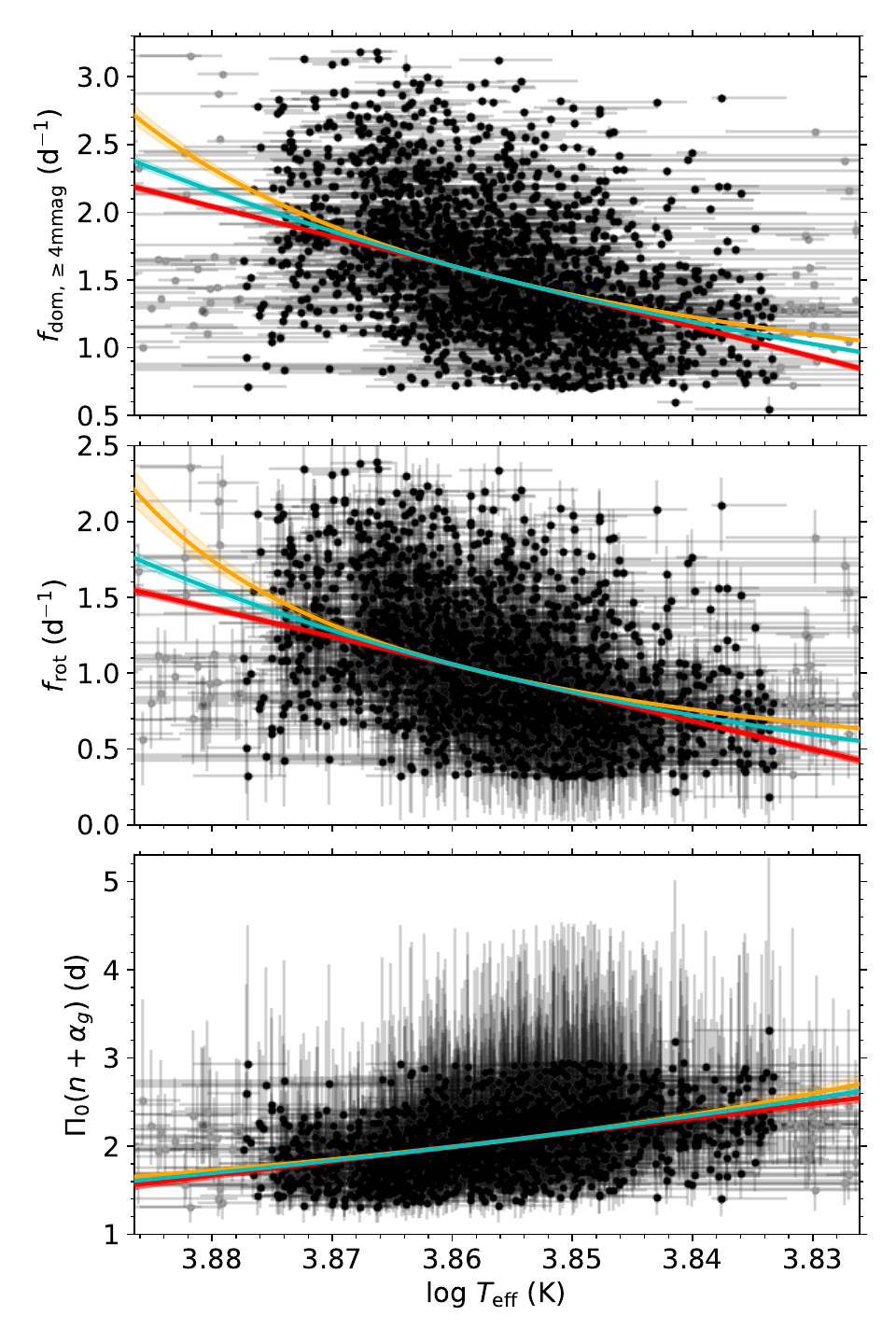}
 \caption{Same as Fig.\,\ref{fig:Gaia_SPB_frot_Teff} but 
   for pulsators within (black) and outside (grey)
   the $\gamma$\,Dor instability region.  The orange, cyan, and red
   lines show the results of inverse, logarithmic, and linear
   regression analyses for the black dots with coefficients listed in
   Table\,\ref{tab:Gaia_frot_Teff}, respectively.
  \label{fig:Gaia_frot_Teff}}
\end{figure}
\section{Extra figures and tables for Sect.\,4}
\begin{figure}[h!]
\hspace{-0.5cm}    \includegraphics[width=10.cm]{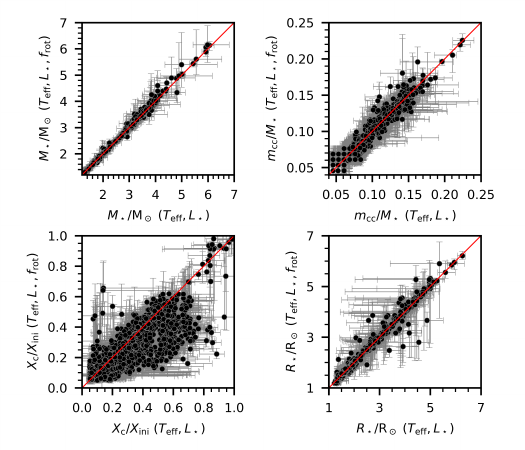}
    \caption{Similar to Fig.\,\ref{fig:modelling}, but for the 2,375
      gravity-mode pulsators modelled via the $[{\rm M/H}] = -0.5$
      grid.}
    \label{fig:modelling-metalpoor}
\end{figure}
\begin{figure}[h!]
\hspace{-0.5cm}    \includegraphics[width=10.cm]{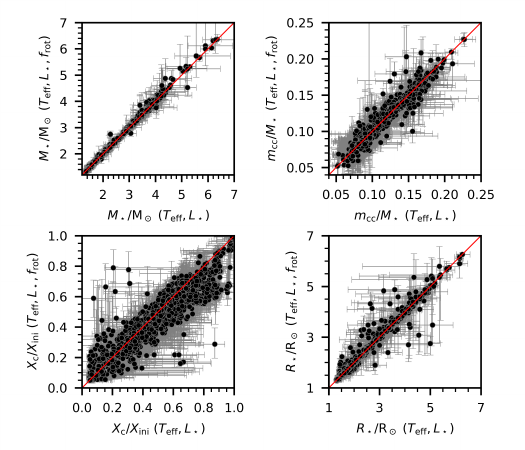}
    \caption{ {Similar to Fig.\,\ref{fig:modelling}, but for the 2,439
        gravity-mode pulsators modelled via the $[{\rm M/H}] = -0.25$
        grid.}}
    \label{fig:modelling-m025}
\end{figure}
\begin{figure*}[h!]
\centering
\rotatebox{270}{\includegraphics[width=6.0cm]{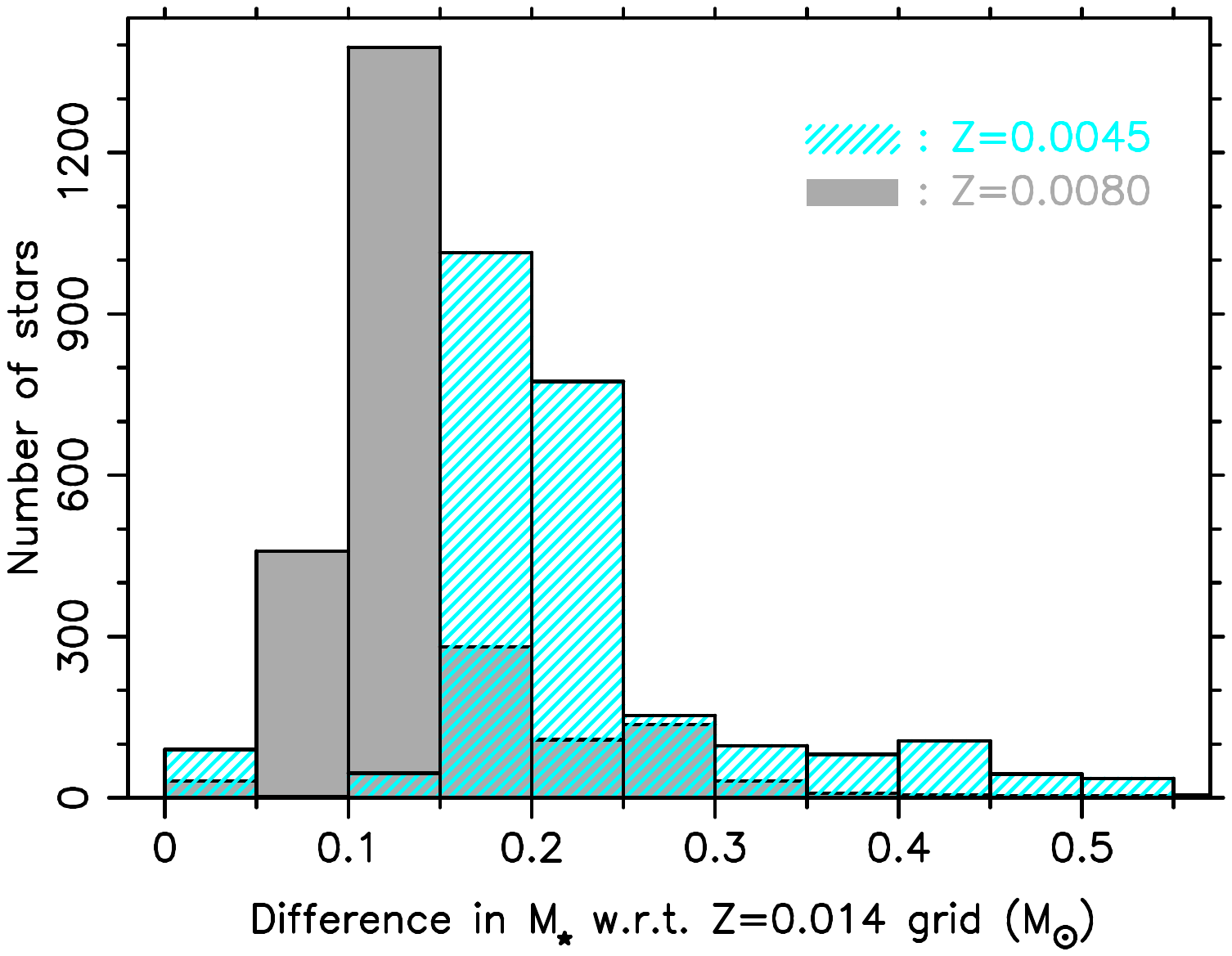}}
\rotatebox{270}{\includegraphics[width=6.0cm]{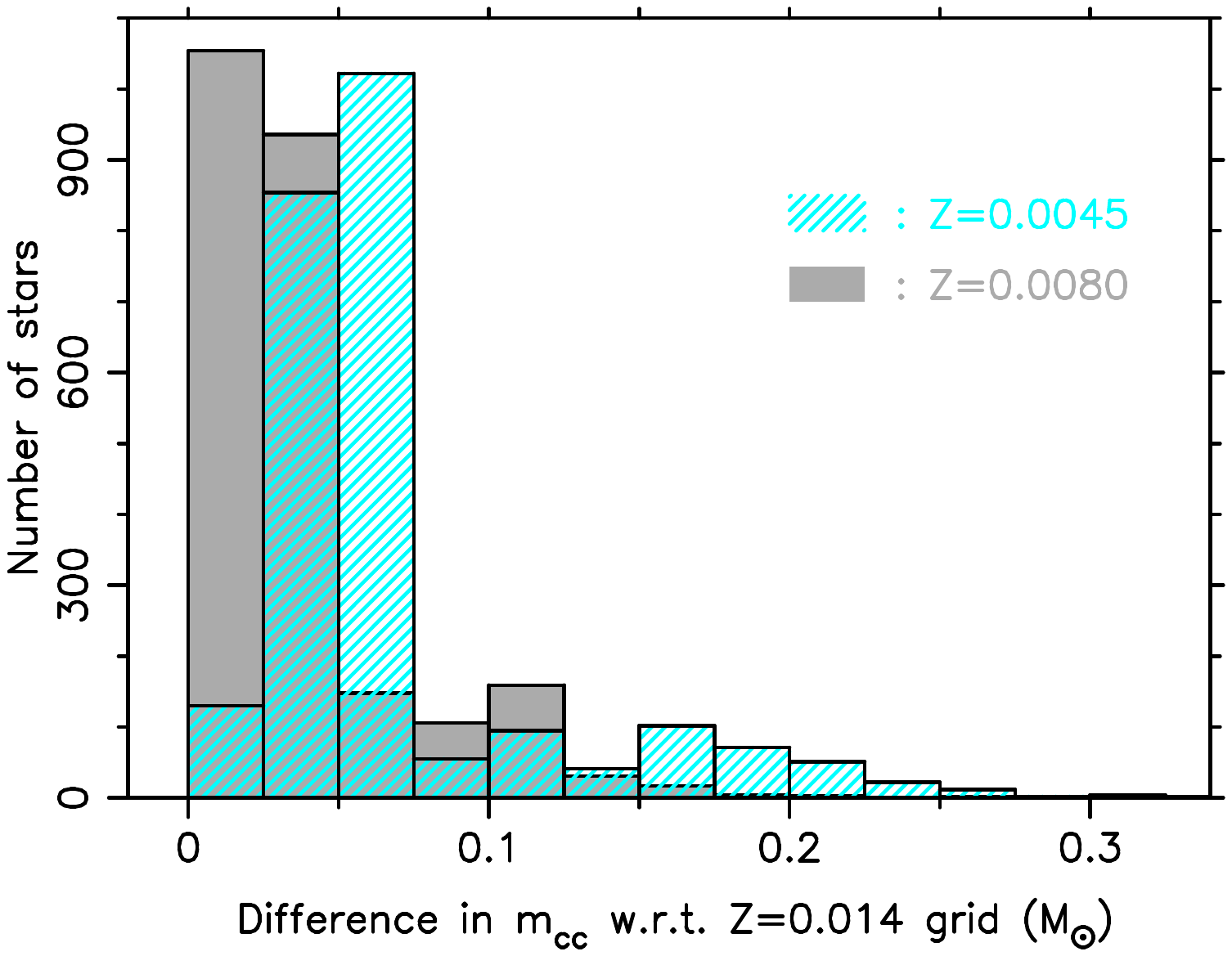}}
\rotatebox{270}{\includegraphics[width=6.0cm]{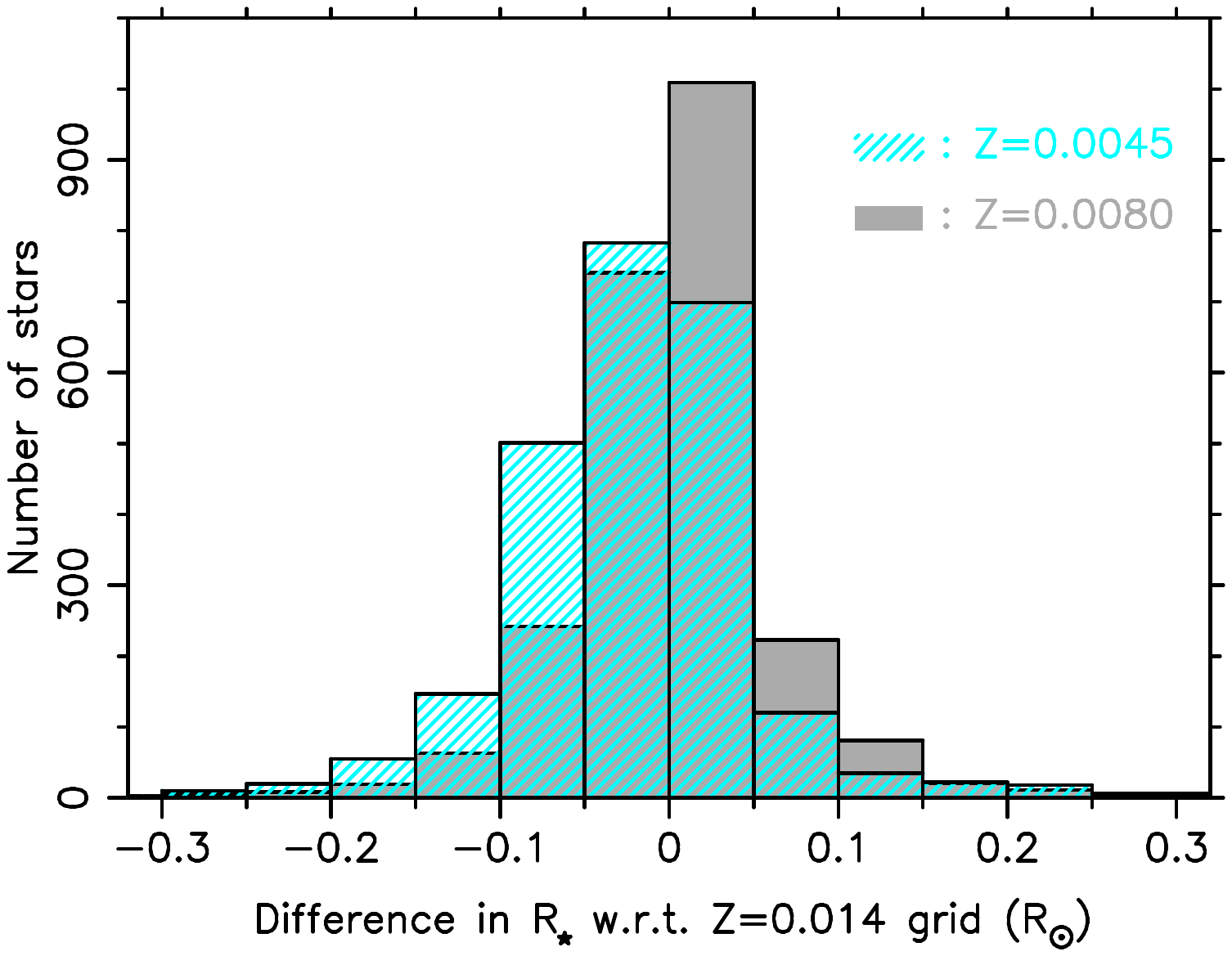}}
\rotatebox{270}{\includegraphics[width=6.0cm]{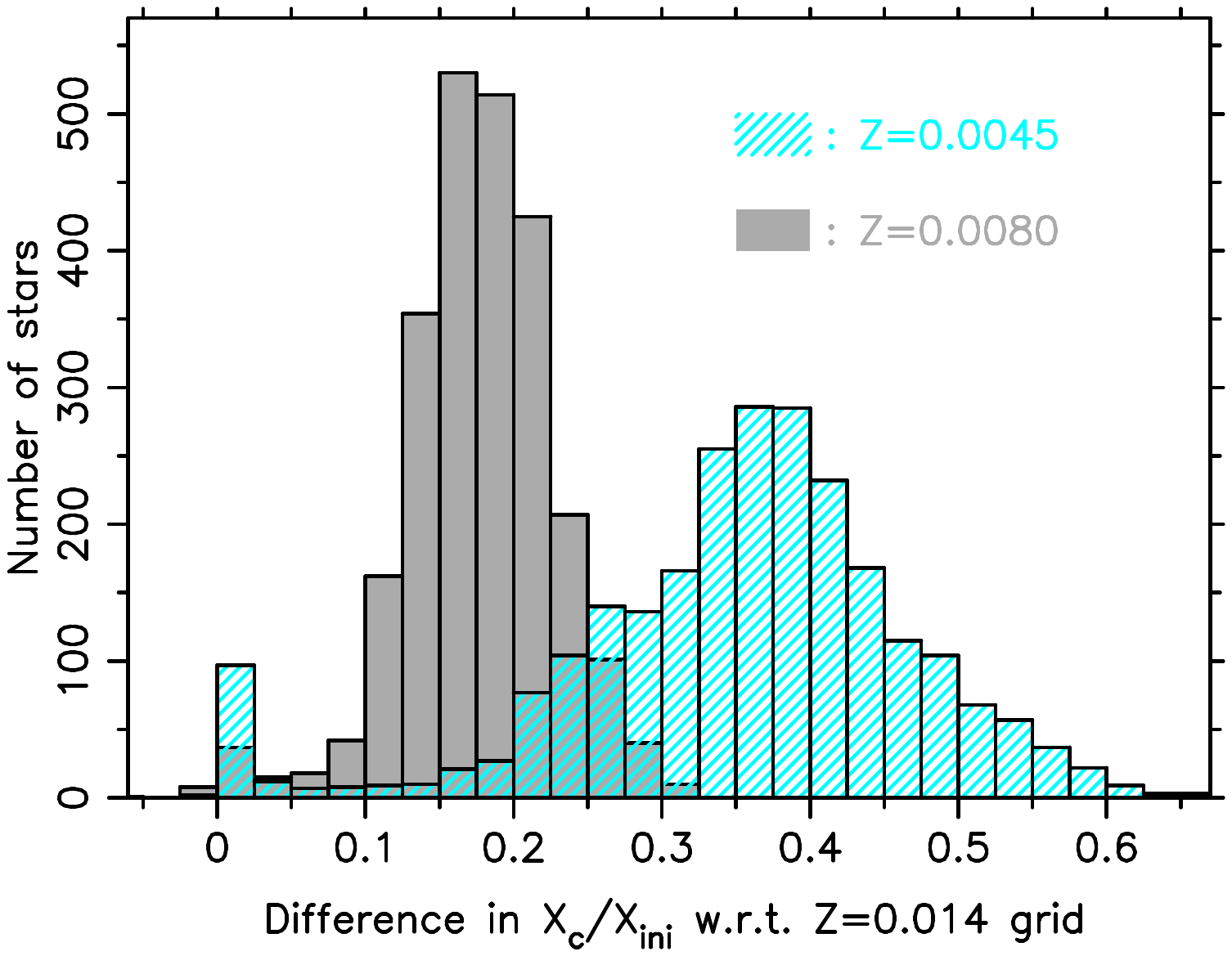}}
\caption{Differences in estimates from the grid modelling for the mass
       (upper left), convective core mass (upper right), radius (lower
       left), and evolutionary stage (lower right) due to the change
       in metallicity. The differences arise from the values for
       $Z=0.014$ minus those for $Z=0.008$ (grey) and $Z=0.0045$
       (cyan), respectively.    }
\label{fig:syshistograms}
\end{figure*}
\begin{figure}[h!]
\hspace{-0.5cm}\rotatebox{0}{\includegraphics[width=9.cm]{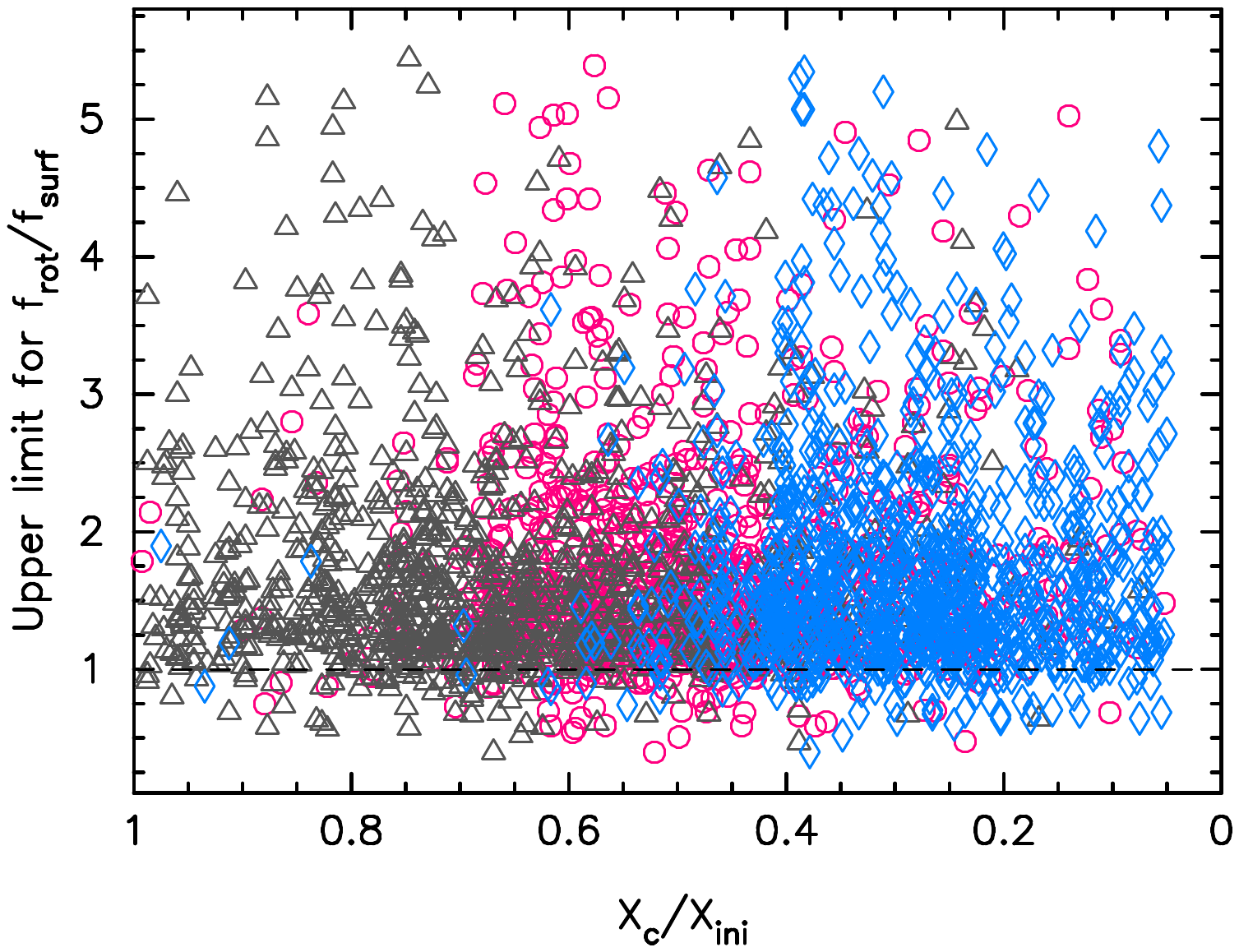}}
\vspace{-0.5cm}
\caption{Upper limits for the near-core to envelope rotation rate
  deduced from {\it Gaia\/} DR3 measurements of {\tt vbroad} available
  for 969, 965, and 929 gravity-mode pulsators in the sample of pulsators
  with a radius estimate from the  $Z=0.014$ (grey triangles),
  $Z=0.0080$ (pink circles), and  $Z=0.0045$ (blue diamonds) model grids, respectively.
\label{fig:upperlimits}}
\end{figure}

\landscape
\begin{table}
\tabcolsep=2pt
    \caption{Parameters of the gravity-mode pulsators covered by the solar metallicity grid.} 
    \label{table-solar}
    {\tiny
    \begin{tabular}{cccccccccccccccccccc}
        \hline
        \hline
        {\it Gaia\/} DR3 ID & 
        $f_{\rm rot}$ & 
        $a_{f_{\rm rot}}$ & 
        $b_{f_{\rm rot}}$ &
        M$_\star$ &
        $a_{{\rm M}_\star}$ & 
        $b_{{\rm M}_\star}$ & 
        $X_{\rm c}/X_{\rm ini}$ & 
        $a_{X_{\rm c}/X_{\rm ini}}$ & 
        $b_{X_{\rm c}/X_{\rm ini}}$ & 
        $m_{\rm cc}$ & 
        $a_{m_{\rm cc}}$ &
        $b_{m_{\rm cc}}$ & 
        $\log (R_\star/R_\odot)$ &
        $a_{\log (R_\star/R_\odot)}$ &
        $b_{\log (R_\star/R_\odot)}$ &
        $f_{\rm co-rot}$ & 
        spin & 
        $f_{\rm surf}\cdot\sin i$ & 
        $f_{\rm rot}/(f_{\rm surf}\cdot\sin i)$ \\
        \hline
164941072177351424 & 0.93767 & 0.16430 & 0.12293 & 3.77256 & 0.07590 & 0.04231 & 0.51880 & 0.06017 & 0.03535 & 0.60150 & 0.04841 & 0.02017 & 0.53383 & 0.01765 & 0.01497 & 0.51010 & 3.67642 & --- & --- \\
 185403705126528256& 1.16134 & 0.16431 & 0.12293 & 1.50733 & 0.01523 & 0.01123 & 0.80702 & 0.04583 & 0.03352 & 0.13033 & 0.01222 & 0.00585 & 0.18797 & 0.01018 & 0.00689 & 0.55408 & 4.19196 & --- & --- \\
\vdots&\vdots&\vdots&\vdots&\vdots&\vdots&\vdots&\vdots&\vdots&\vdots&\vdots&\vdots&\vdots&
 \vdots&\vdots&\vdots&\vdots&\vdots&\vdots&\vdots\\
 200619262425969024 & 1.64651 & 0.16497 & 0.12382 & 1.48797 & 0.03052 & 0.01000 & 0.89223 & 0.07543 & 0.04644 & 0.12030 & 0.02328 & 0.00319 & 0.17293 & 0.01266 & 0.01166 & 0.64949 & 5.07016 & 1.42700 & 1.15383      \\ \vdots&\vdots&\vdots&\vdots&\vdots&\vdots&\vdots&\vdots&\vdots&\vdots&\vdots&\vdots&\vdots&
 \vdots&\vdots&\vdots&\vdots&\vdots&\vdots&\vdots\\
 \hline\end{tabular}\\[-0.2cm]
    }
\tablefoot{When available, the error estimates of a quantity $X$ are
  given as $a_X$ and $b_X$ in the notation
  $X_{-a_X}^{+b_X}$. Frequencies are given in the unit d$^{-1}$ while
  the stellar (core) mass and radius are given in solar units. The
  quantity $f_{\rm co-rot}$ is the frequency of the dominant
  gravito-inertial mode in a frame of reference co-rotating with
  $f_{\rm rot}$, while `spin' stands for the spin parameter of the
  dominant mode in the region just outside the convective core.} \\[1cm] 
\end{table}

\begin{table}
\tabcolsep=2pt
    \caption{Same as Table\,\ref{table-solar} but for the grid having
    $Z=0.0080$. }
    \label{table-Z0-008}
    {\tiny
    \begin{tabular}{cccccccccccccccccccc}
        \hline
        \hline
        {\it Gaia\/} DR3 ID & 
        $f_{\rm rot}$ & 
        $a_{f_{\rm rot}}$ & 
        $b_{f_{\rm rot}}$ &
        M$_\star$ &
        $a_{{\rm M}_\star}$ & 
        $b_{{\rm M}_\star}$ & 
        $X_{\rm c}/X_{\rm ini}$ & 
        $a_{X_{\rm c}/X_{\rm ini}}$ & 
        $b_{X_{\rm c}/X_{\rm ini}}$ & 
        $m_{\rm cc}$ & 
        $a_{m_{\rm cc}}$ &
        $b_{m_{\rm cc}}$ & 
        $\log (R_\star/R_\odot)$ &
        $a_{\log (R_\star/R_\odot)}$ &
        $b_{\log (R_\star/R_\odot)}$ &
        $f_{\rm co-rot}$ & 
        spin & 
        $f_{\rm surf}\cdot\sin i$ & 
        $f_{\rm rot}/(f_{\rm surf}\cdot\sin i)$ \\
        \hline
         164941072177351424 & 
         0.93767 & 0.16430 & 0.12293 & 3.52086 & 0.07890 & 0.05629 & 
         0.37845 & 0.05959 & 0.04983 & 0.48120 & 0.04536 & 0.03724 &
         0.53008 & 0.01471 & 0.01843 & 0.51010 & 3.67642 & --- & --- \\
 185403705126528256 & 1.16134 & 0.16431 & 0.12293 & 1.37180 & 0.01456 & 0.02682 & 0.60652 & 0.05884 & 0.02316 & 0.12030 & 0.01738 & 0.00291 & 0.18797 & 0.00813 & 0.01017 & 0.55408 & 4.19196 & --- & --- \\
\vdots&\vdots&\vdots&\vdots&\vdots&\vdots&\vdots&\vdots&\vdots&\vdots&\vdots&\vdots&\vdots&
 \vdots&\vdots&\vdots&\vdots&\vdots&\vdots&\vdots\\
200619262425969024 &  1.64651 & 0.16497 & 0.12382 & 1.35244 & 0.01544 & 0.01203 & 0.62155 & 0.04435 & 0.06126 & 0.10025 & 0.01923 & 0.00611 & 
0.17669 & 0.01827 & 0.01008 & 0.64949 & 5.07016 & 1.41470 & 1.16386 \\
\vdots&\vdots&\vdots&\vdots&\vdots&\vdots&\vdots&\vdots&\vdots&\vdots&\vdots&\vdots&\vdots&
 \vdots&\vdots&\vdots&\vdots&\vdots&\vdots&\vdots\\
 \hline\end{tabular}\\[1cm]
 }
\end{table}

\begin{table}
\tabcolsep=2pt
    \caption{Same as Table\,\ref{table-solar} but for the grid having
    $Z=0.0045$. }
    \label{table-Z0-0045}
    {\tiny
    \begin{tabular}{cccccccccccccccccccc}
        \hline
        \hline
        {\it Gaia\/} DR3 ID & 
        $f_{\rm rot}$ & 
        $a_{f_{\rm rot}}$ & 
        $b_{f_{\rm rot}}$ &
        M$_\star$ &
        $a_{{\rm M}_\star}$ & 
        $b_{{\rm M}_\star}$ & 
        $X_{\rm c}/X_{\rm ini}$ & 
        $a_{X_{\rm c}/X_{\rm ini}}$ & 
        $b_{X_{\rm c}/X_{\rm ini}}$ & 
        $m_{\rm cc}$ & 
        $a_{m_{\rm cc}}$ &
        $b_{m_{\rm cc}}$ & 
        $\log (R_\star/R_\odot)$ &
        $a_{\log (R_\star/R_\odot)}$ &
        $b_{\log (R_\star/R_\odot)}$ &
        $f_{\rm co-rot}$ & 
        spin & 
        $f_{\rm surf}\cdot\sin i$ & 
        $f_{\rm rot}/(f_{\rm surf}\cdot\sin i)$ \\
        \hline
         164941072177351424 & 0.93767 & 0.16430 & 0.12293 & 3.34662 & 0.07221 & 0.05528 & 0.28070 & 0.09347 & 0.03254 & 0.42105 & 0.06547 & 0.01727 & 0.54135 & 0.01826 & 0.01336 & 0.51010 & 3.67642 & --- & --- \\
 185403705126528256 & 1.16134 & 0.16431 & 0.12293 & 1.31372 & 0.01894 & 0.00624 & 0.37343 & 0.06841 & 0.03501 & 0.09023 & 0.00987 & 0.00860 & 0.19549 & 0.01133 & 0.00620 & 0.55408 & 4.19196 & --- & --- \\
\vdots&\vdots&\vdots&\vdots&\vdots&\vdots&\vdots&\vdots&\vdots&\vdots&\vdots&\vdots&\vdots&
 \vdots&\vdots&\vdots&\vdots&\vdots&\vdots&\vdots\\
200619262425969024 &  1.64651 & 0.16497 & 0.12382 & 1.31372 & 0.01858 & 0.01000 & 0.34837 & 0.04428 & 0.06441 & 0.09023 & 0.01349 & 0.00586 & 0.19925 & 0.00852 & 0.01219 & 0.64949 & 5.07016 & 1.34308 & 1.22592 \\ 
\vdots&\vdots&\vdots&\vdots&\vdots&\vdots&\vdots&\vdots&\vdots&\vdots&\vdots&\vdots&\vdots&
 \vdots&\vdots&\vdots&\vdots&\vdots&\vdots&\vdots\\
 \hline\end{tabular}
 }
\end{table}

\end{document}